\documentclass[12pt,a4paper]{article}
\usepackage{graphicx}
\usepackage{amssymb}
\usepackage{amsmath}
\usepackage{bm}
\usepackage{color}
\usepackage{tikz}
\usepackage{theorem}
\usepackage{cite}
\usepackage{amsfonts}

\usepackage{graphicx}
\usepackage{caption}
\usepackage{subcaption}

\usepackage[sort&compress,numbers, merge]{natbib}

\newcommand{\eprint}[2][]{\href{https://arxiv.org/abs/#2}{arXiv:~\nolinkurl{#2}}}

\definecolor{Orange}{cmyk}{0,0.61,0.87,0}
\definecolor{JungleGreen}{cmyk}{0.99,0,0.52,0} 
\definecolor{OliveGreen}{cmyk}{0.64,0,0.95,0.40}
\definecolor{Brown}{cmyk}{0,0.81,1,0.60}
\definecolor{RoyalBlue}{cmyk}{0.71,0.53,0,0.12}
\definecolor{darkspringgreen}{rgb}{0.09, 0.45, 0.27}

\newcommand{\be}{\begin{equation}}
\newcommand{\ee}{\end{equation}}
\newcommand{\bea}{\begin{eqnarray}}
\newcommand{\eea}{\end{eqnarray}}
\newcommand{\eq}[1]{Eq.~(\ref{#1})}
\newcommand{\la}{\big\langle}
\newcommand{\ra}{\big\rangle}

\newcommand{\M}{{\cal M}}

\usetikzlibrary{decorations.markings}
\usetikzlibrary{fadings}
\usetikzlibrary{shapes.misc}
\tikzset{crossr/.style={cross out, draw=red, minimum size=4*(#1-\pgflinewidth), inner sep=0pt, outer sep=0pt},
crossr/.default={2pt}}
\tikzset{crossb/.style={cross out, draw=black, minimum size=4*(#1-\pgflinewidth), inner sep=0pt, outer sep=0pt},
crossb/.default={2pt}}
\tikzset{crossp/.style={cross out, draw=violet, minimum size=4*(#1-\pgflinewidth), inner sep=0pt, outer sep=0pt},
crossp/.default={2pt}}

\tikzfading 
[
  name=fade out,
  inner color=transparent!0,
  outer color=transparent!100
]

\usepackage[colorlinks=True, citecolor=blue, linkcolor=blue, urlcolor=black]{hyperref}

\def\nmax{n_{\rm max}}

\allowdisplaybreaks[1]

\begin{document}

\begin{titlepage}
\begin{center}
\hfill

\vspace{2.0cm}
{\Large\bf  
Cornering Large-$N_c$ QCD with Positivity Bounds}

\vspace{2cm}
{\bf 
Clara Fernandez$^1$, Alex Pomarol$^{1,2,3}$, Francesco Riva$^4$ and Francesco Sciotti$^1$
}
\\
\vspace{0.7cm}
{\it\footnotesize
${}^1$IFAE and BIST, Universitat Aut\`onoma de Barcelona, 08193 Bellaterra, Barcelona\\
${}^2$Departament de F\'isica, Universitat Aut\`onoma de Barcelona, 08193 Bellaterra, Barcelona\\
${}^3$CERN, Theory Division, Geneva, Switzerland\\
${}^4$D\'epartment de Physique Th\'eorique, Universit\'e de Gen\`eve,
24 quai Ernest-Ansermet, 1211 Gen\`eve 4, Switzerland
}

\vspace{0.9cm}
\abstract

The simple  analytic structure of meson scattering amplitudes in the large-$N_c$ limit, combined with
 positivity of the spectral density, provides precise  predictions on low-energy observables.
Building upon previous studies, we explore the allowed regions of  chiral Lagrangian parameters  and meson couplings to pions.
We reveal a structure of kinks at all orders in the chiral expansion and  develop analytical tools to show that  kinks always correspond to  amplitudes with a single light pole.
We build (scalar- and vector-less) deformations of the Lovelace-Shapiro and Coon UV-complete amplitudes, and show that they
lie close  to the boundaries. 
Moreover,  constraints from crossing-symmetry imply that meson couplings to pions become smaller as their spin increases, providing an explanation for the success of Vector Meson Dominance and holographic QCD.
We study how these conclusions depend on assumptions about the high-energy behavior of amplitudes.
Finally, we  emphasize the complementarity between our results and  Lattice computations  
in the exploration of large-$N_c$~QCD.
\end{center}
\end{titlepage}
\setcounter{footnote}{0}

\section{Introduction}

The understanding of theories at strong coupling  is one of the most important challenges
of modern particle physics. Besides the pragmatic interest  for QCD hadronic physics,
 such understanding would  broaden our perspective on  theories  beyond the Standard Model and plausibly provide intuition on its shortcomings.

Dispersion relations distill the essential ingredients of quantum field theory, unitarity and causality, into consistency conditions for scattering amplitudes. They have been used as \emph{positivity bounds} to shape the parameter space of  effective field theories originating from healthy, albeit strongly coupled, microscopic dynamics, see e.g. \cite{Adams:2006sv,Arkani-Hamed:2020blm,deRham:2017avq,Bellazzini:2020cot,Tolley:2020gtv,Caron-Huot:2020cmc,Caron-Huot:2021rmr}, and have had important applications to QCD and the chiral Lagrangian \cite{Pham:1985cr,Ananthanarayan:1994hf,Pennington:1994kc,Bijnens:1997vq,Ananthanarayan:2000ht,Colangelo:2001df,Manohar:2008tc,Guerrieri:2019rwp,Guerrieri:2020bto}.
 
At the same time, in the context of $SU(N_c)$ gauge theories,  the limit of many colors $N_c\to \infty$ 
has provided one of the most insightful  approaches for understanding the strongly-coupled regime~\cite{tHooft:1973alw,Witten:1979kh}, even though  real-world QCD has only $N_c=3$.
The main consequence  of this  approximation is that the theory  has a dual description 
in terms of weakly-coupled  mesons, rather than quarks and gluons.  
In spite of this important step forward,
the predictions of  large-$N_c$ QCD have been limited by the fact  that the theory contains an  infinite number of mesons  of any spin, whose Lagrangian is unknown or contains an infinite number of terms. 

Recently,  Ref.~\cite{Albert:2022oes} has combined these two approaches and  derived important constraints on the low energy $\pi\pi\to \pi\pi$ scattering amplitude.
{Despite  the  higher-spin  meson spectrum  remains unknown,
 the simple analytic structure of   large-$N_c$ amplitudes -- combined with certain assumptions 
 on their high-energy behavior -- improves the predictiveness  of dispersion relations.}

In this article we push forward this approach, combining analytic and numerical methods. 
One of the most important  questions raised in Ref.~\cite{Albert:2022oes} 
concerns the understanding of 
which theories define
the kinks (and the bulk) of the allowed regions of  Wilson coefficients.
In this article we identify these theories. 
{We build  UV-complete  four-pions  amplitudes, as variations of 
the Lovelace-Shapiro amplitude \cite{Lovelace:1968kjy,Shapiro:1969km}
 and the Coon amplitude \cite{Coon:1972qz} in which the spin-0 (and spin-1) poles have been removed, and show that they reside close to  the boundary.} At the kinks, on the other hand, we find  theories  with an infinitely degenerate higher-spin spectrum, or theories with a unique state {at finite mass}, either of spin
$J=0$, $J=1$ or  $J=2$. While some of these results are known \cite{Bellazzini:2020cot,Caron-Huot:2020cmc,Albert:2022oes}, in this work we are able to exclude  other possibilities. 
In particular, while the numerical approach  suggests the existence of a new kink with a more exotic spectrum, we show  that the kink position can be reformulated as a 1D moment-problem whose solution  converges to the $J=1$ theory.

An important emerging property of low-energy QCD is
Vector Meson Dominance (VMD)~\cite{Sakurai},  the hypothesis that the spin-1 $\rho$ meson
gives the main contribution to the chiral Lagrangian~\cite{Ecker:1988te}.
This property has an  empirical origin (it works!),  but it lacks a  theoretical explanation.
We will show that VMD  finds   its  origin in the positivity bounds. Indeed,
 the  chiral Lagrangian coefficients can be written  
as a  sum over positive quantities that depend on the couplings and masses of the different mesons:
when  the theory contains a  $\rho$,  this    dominates  over the other $J>1$ states. 
This is especially prominent when the higher-spin states are heavier than the $\rho$, as it occurs in real-world QCD.  
Our analysis also  provides insight into why holographic models, which contain only (charged) spin-0 and spin-1 states, have been so successful in predicting low-energy properties of QCD, see e.g. \cite{Erlich:2005qh,DaRold:2005mxj}.
Finally, we will show that this reasoning also extends to the heavy-meson couplings to pions: the higher the spin, the smaller the coupling.

These arguments, as well as all those from Ref.~\cite{Albert:2022oes}, rely on assuming that the high-energy amplitudes are particularly well behaved, ${\cal M}/s\to 0$ at large $|s|$. In this article we will also discuss the implications of relaxing this assumption, such that the $\pi\pi\to\pi\pi$ amplitude is limited only by the Froissart-Martin bound ${\cal M}/s^2\to 0$
at large energies~\cite{Froissart:1961ux,Martin:1962rt}.
 In this case, the spin-1 and spin-0 contributions decouple from (and can contribute more than) the 
$J\geq2$ ones, that are now dominated by the spin-2 state instead, in a generalisation of VMD.

The paper is organised as follows. 
In Sec.~\ref{sec2} we review the analytical structure of the 
$\pi \pi \rightarrow \pi \pi$ amplitude and  the dispersion relations that lead to  positivity constraints.
We also present possible UV completions to the chiral Lagrangian, as they will play an important role
 to understand  the boundaries of the Wilson-coefficient allowed regions. 
In Sec.~\ref{ms} we consider the case in which the four-pion amplitude  at large $|s|$ satisfies
${\cal M}/s\to 0$. We determine the allowed regions of the leading Wilson coefficients
and show which theories reside at the kinks of these boundaries.
We also study the emergence of VMD and derive bounds on the couplings of meson resonances to pions.
In Sec.~\ref{Ms2} we extend the analysis to the case  in which 
the four-pion amplitude only satisfies the Froissart-Martin bound at high energies.
We present several appendices with extended discussions on the numerical bootstrap (Appendix~\ref{appa}),
on the  analytical determination of the kinks (Appendix~\ref{appb}), the $su$-models (Appendix~\ref{appc}), and 
the Lovelace-Shapiro and Coon amplitudes (Appendix~\ref{appd} and Appendix~\ref{appe} respectively).

\section{The $\pi \pi \rightarrow \pi \pi$ Amplitude in large-$N_c$ QCD}
\label{sec2}
This section contains mostly a review of previous literature on the $2\rightarrow 2 $ pion amplitude,  in particular the results of Ref.~\cite{Albert:2022oes}. 
We will work in the massless quark limit.

Pions  are the massless Goldstone bosons associated to the spontaneous breaking  of  the global $SU(N_f)_L\times SU(N_f)_R\to  SU(N_f)$, where $N_f$ is the number of quark flavors in QCD.
 They transform in the $Adj.$ representation of $SU(N_f)$, which for the case $N_f=2$ corresponds to the Isospin  $I=1$, the  triplet $\pi^\pm$ and $\pi^0$. 
 This allows us to write the $2\rightarrow 2$ pion amplitude as
\be
{\cal M}(\pi^a\pi^b\to \pi^c\pi^d)=A(s|t,u)\left[\frac{2}{N_f}\delta_s+d_s\right]
+A(t|u,s)\left[\frac{2}{N_f}\delta_t+d_t\right]
+A(u|s,t)\left[\frac{2}{N_f}\delta_u +d_u\right]\,,
\ee
where $s=(p_a+p_b)^2$, $t=(p_a-p_c)^2$, $u=(p_a-p_d)^2$, and 
\bea
&& \delta_s=\delta_{ab}\delta_{cd}\ ,\ \delta_t=\delta_s(b\leftrightarrow c)\ , \ \delta_u=\delta_s(b\leftrightarrow d)\,, \\
&&d_s=d_{abe}d_{cde}\ , \ d_t=d_s(b\leftrightarrow c)\ , \ d_u=d_s(b\leftrightarrow d)\,,
\eea
correspond to the various ways of contracting $SU(N_f)$ adjoint indices into singlets.
$A(s|t,u)$~is a function of $t,u$ symmetric under their interchange, i.e. $A(s|t,u)=A(s|u,t)$.

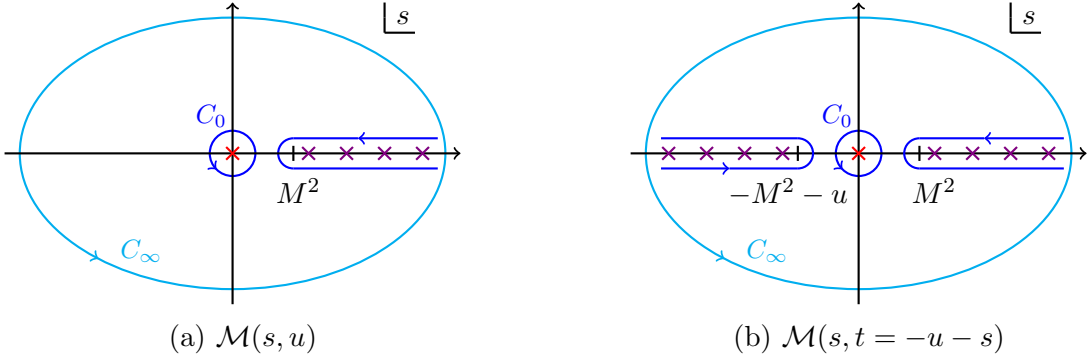
\begin{figure}[t]\hspace{0.1cm}
  \begin{subfigure}[b]{0.5\textwidth}
\begin{tikzpicture}
\begin{scope}[thick,font=\small]
\hskip1.1cm 
\draw[decoration={markings, mark=at position 0.625 with {\arrow{>}}},postaction={decorate}, color=cyan] (3,0) ellipse (2.8 and 1.8) {};
\draw[decoration={markings, mark=at position 0.625 with {\arrow{>}}},postaction={decorate}, color=blue] (3,0) circle (0.3) {};
\node[color=cyan] at (1.8,-1.3) {\footnotesize{$C_\infty$}};
\node[color=blue] at (2.72,0.5) {\footnotesize{$C_0$}};
\draw[color=blue] (3.8,0.2)--(4.65,0.2) {};
\draw[color=blue,<-] (4.65,0.2)--(5.7,0.2) {};
\draw[color=blue] (3.8,-0.2)--(5.7,-0.2) {};
\draw[color=blue] (3.8,0.2) arc (90:270:0.2) {};
\draw[black, thick,->] (0,0) -- (6,0);
\draw[black, thick,->] (3,-2) -- (3,2);
\draw[black] (5,2) -- (5,1.8) node[right] {$s$};
\draw[black] (5,1.8) -- (5,1.6) {};
\draw[black] (5,1.6) -- (5.4,1.6) {};
\draw (3,0) node[crossr] {};
\draw (4,0) node[crossp] {};
\draw[black]  (3.8,0.1) -- (3.8,-0.1) node[midway,below=0.18cm,align=center] {$\ M^2$};
\draw (4.5,0) node[crossp] {};
\draw (5,0) node[crossp] {};
\draw (5.5,0) node[crossp] {};
\end{scope}
\end{tikzpicture}
     \caption{
  ${\cal M}(s,u)$ 
  }
    \label{fig:uchannel}
  \end{subfigure}\hspace{-0.4cm}
  \begin{subfigure}[b]{0.5\textwidth}
\begin{tikzpicture}
\begin{scope}[thick,font=\small]
\hskip1.1cm 
\draw[decoration={markings, mark=at position 0.625 with {\arrow{>}}},postaction={decorate}, color=cyan] (3,0) ellipse (2.8 and 1.8) {};
\draw[decoration={markings, mark=at position 0.625 with {\arrow{>}}},postaction={decorate}, color=blue] (3,0) circle (0.3) {};
\node[color=cyan] at (1.8,-1.3) {\footnotesize{$C_\infty$}};
\node[color=blue] at (2.72,0.5) {\footnotesize{$C_0$}};
\draw[black] (5,2) -- (5,1.8) node[right] {$s$};
\draw[black] (5,1.8) -- (5,1.6) {};
\draw[black] (5,1.6) -- (5.4,1.6) {};
\draw[black, thick,->] (0,0) -- (6,0);
\draw[black, thick,->] (3,-2) -- (3,2);
\draw (3,0) node[crossr] {};
\draw (4,0) node[crossp] {};
\draw[black]  (3.8,0.1) -- (3.8,-0.1) node[midway,below=0.18cm,align=center] {$\ \ \  M^2$};
\draw[black]  (2.2,0.1) -- (2.2,-0.1) node[midway,below=0.18cm,align=center] {$-M^2-u \ \ $};
\draw (4.5,0) node[crossp] {};
\draw (5,0) node[crossp] {};
\draw (5.5,0) node[crossp] {};
\draw (0.5,0) node[crossp] {};
\draw (1,0) node[crossp] {};
\draw (1.5,0) node[crossp] {};
\draw (2,0) node[crossp] {};
\draw[color=blue] (3.8,0.2)--(4.65,0.2) {};
\draw[color=blue,<-] (4.65,0.2)--(5.7,0.2) {};
\draw[color=blue] (3.8,-0.2)--(5.7,-0.2) {};
\draw[color=blue] (3.8,0.2) arc (90:270:0.2) {};
\draw[color=blue] (0.4,0.2)--(2.2,0.2) {};
\draw[color=blue,->] (0.4,-0.2)--(1.3,-0.2) {};
\draw[color=blue] (1.3,-0.2)--(2.2,-0.2) {};
\draw[color=blue] (2.2,0.2) arc (90:-90:0.2) {};
\end{scope}
\end{tikzpicture}
    \caption{${\cal M}(s,t=-u-s)$   
    }
    \label{fig:tchannel}
  \end{subfigure}
  \caption{\it Analytic  structure of  ${\cal M}(s,u)$  and  ${\cal M}(s,t=-s-u)$ for fixed $u<0$. We denote by $C_0$,  $C_\infty$ (to be taken at $|s|\to \infty$) and the discontinuity along the real axis  the relevant contours of integration used for  dispersion relations.}
    \label{fig:sumrules}
\end{figure}

In the large-$N_c$ limit, QCD reduces to a theory of weakly coupled mesons,  whose trilinear couplings scale as $\sim 1/\sqrt{N_c}$  \cite{tHooft:1973alw,Witten:1979kh}.  In this limit,  the $2\rightarrow 2$ pion amplitude
is then dominated by a tree-level meson exchange. 
Since these mesons are $q\bar q$ states with isospin $I=0,1$, the isospin $I=2$ amplitude, 
\be
{\cal M}^{I=2}_s(\pi^a\pi^b\to \pi^c\pi^d)= A(t|u,s)+A(u|s,t)\equiv {\cal M}(t,u)\,,
\label{i2schannel}
\ee 
(symmetric under $t\leftrightarrow u$) {has no poles in the large-$N_c$ limit}.
All this leads to the following important implications for the analytical structure of ${\cal M}$ in the large-$N_c$ limit: 
\begin{itemize}
\item The only singularities of  ${\cal M}$  in the complex $s$-plane are simple poles associated with the tree-level meson exchange (the branch cut along the physical region is at least $O(1/N_c^2$)).
\item The absence of $I=2$ meson exchange in the $s$-channel implies that ${\cal M}_s^{I=2}={\cal M}(t,u)$ has no poles for real $s>0$. Since $t=-s-u$, this implies that for fixed $u<0$, there cannot be poles in $\M(t,u)$ on
the  negative real $t$ axis. 
Now, by a simple  exchange of  arguments ($t\to s$),
we come to the conclusion that ${\cal M}(s,u)$, for fixed $u<0$, can only have poles on the positive real $s$ axis. 
\item For fixed $u<0$, 
 ${\cal M}(s,t)$  can have  poles either on the real positive or negative $s$ axis.
\end{itemize}
This analytic structure of ${\cal M}$ is  illustrated in Fig.~\ref{fig:sumrules}.

At energies below the mass $M$ of the lightest massive meson (corresponding to the position of the first pole in Fig.~\ref{fig:sumrules}), pions can be well described by an effective theory, corresponding to an expansion in $s/M,u/M\to 0$, 
\bea
{\cal M}(s,u)
&=& \sum_{n=1}^\infty\sum_{l=0}^{[n/2]} \, g_{n,l}\,  s^{\{n-l}  u^{l\}}\nonumber\\
&=& g_{1,0}\,(s+u)+g_{2,0}\,(s^2+u^2)+ g_{2,1} su+g_{3,0}\,(s^3+u^3)\nonumber\,
 \\
 &+&  g_{3,1}\,(s^2 u+s u^2)+g_{4,0}\,(s^4+u^4)+g_{4,1}\,(s^3 u+s u^3)+ g_{4,2}\, s^2 u^2+...\,,
\label{chila}
\eea
where for convenience we have defined the weighed symmetric tensor $s^{\{i}  u^{j\}}\equiv s^iu^j+(1-\delta_{ij})s^ju^i$ to avoid double counting of terms with $n=2l$, such as $g_{2,1}$, $g_{4,2}$, etc.
A constant term is absent in \eq{chila} as the amplitude must go to zero for $s,u\to 0$ in order to restore the Adler condition for pions.  In the large-$N_c$ limit all Wilson coefficients scale as $g_{n,l}\sim 1/N_c$.
For the connection of \eq{chila} with the QCD chiral Lagrangian, see Sec.~\ref{QCDcon}.


 \subsection{Dispersion Relations and Sum Rules}\label{sec:disprel}

Dispersion relations can be derived assuming that the amplitude ${\cal M}$ satisfies the following high-energy conditions, for fixed $u<0$ and for all $k\geq{k_{\rm min}}$,

\vspace{5mm}
\begin{subequations}\label{UVeq}
\begin{minipage}{0.4\textwidth}
\begin{equation}
\lim_{|s|\to\infty}\frac{{\cal M}(s,u)}{s^k} \to 0\,, \label{UVa}
\end{equation}
  \end{minipage}%
  \hspace{0.05\textwidth}
  \begin{minipage}{0.4\textwidth}
  \vspace{-2mm}
\begin{equation}
\lim_{|s|\to\infty}\frac{{\cal M}(s,-u-s)}{s^k} \to 0\, . \label{UVb}
\end{equation}
    \end{minipage}
\end{subequations}
 \vspace{5mm}
 
 \noindent
Different assumptions about the high-energy behavior of amplitudes are associated with different values of~${k_{\rm min}}$.
On general grounds, the Froissart-Martin bound \cite{Froissart:1961ux,Martin:1962rt,Jin:1964zza} ensures that  in theories with a mass gap Eqs.~(\ref{UVeq}) are satisfied for $k_{\rm min}=2$, with similar results for the massless case~\cite{Caron-Huot:2021rmr,Haring:2022cyf,Herrero-Valea:2020wxz}.

On the other hand, Ref.~\cite{Albert:2022oes}  advocated -- invoking  considerations on the Pomeron Regge trajectory  --  that the four-pion amplitude might be bounded by $\M\lesssim s$ and therefore $k_{\rm min}=1$. In this article  we will study both,  the case with ${k_{\rm min}}=1$ and ${k_{\rm min}}=2$.

\subsubsection{IR-UV Relations}
Taking  $u<0$ fixed, we have that the integral of ${\cal M}(s,u)/s^{k+1}$ 
 along the contour $C_\infty$ of Fig.~\ref{fig:uchannel}   vanishes for $k\geq{k_{\rm min}}$, due to \eq{UVa}.
Because of amplitude's analyticity, we can deform $C_\infty$ into  the blue contour in Fig.~\ref{fig:uchannel}, 
\be
\oint_{\rm C_0} ds'\ \frac{{\cal M}(s',u)}{s^{\prime k+1}}= 2 i
\int_{M^2}^\infty  ds'\ \frac{{\rm Im} {\cal M}(s',u)}{s^{\prime k+1}}\,.
\label{SU}
\ee
The amplitude can be expanded in partial waves,
\bea
{\rm Im} {\cal M}(s,u) = \sum_J (2J+1)\rho_J(s) P_J\bigg( 1 + \frac{2u}{s}\bigg)\,,
\label{pw}
\eea
where $P_J$ are the Legendre polynomials and $\rho_J(s)$, the spectral density, must be positive,
  $\rho_J(s)\geq0$, due to unitarity of the S-Matrix.
  Indeed, for large-$N_c$ theories, we have that the spectral density is given by
\be\label{weakspectral}
(2J+1)\rho_J(m^2)=\pi \sum_i {g^2_{i\pi\pi}} m^2_i\,\delta(m^2-m^2_i)\delta_{JJ_i}\,,
\ee
where $i$ labels mesons of mass $m_i$, spin $J_i$ and coupling to pions $g_{i\pi\pi}$.
  
 Plugging \eq{pw} into \eq{SU},  performing the contour integrals,
and  expanding around small  $u<0$ we find
\bea
k=1: \qquad g_{1,0}+g_{2,1} u+g_{3,1}u^2+... &=&\left\langle \frac{P_J(1)}{m^2}+2\frac{P'_J(1)}{m^4}u+2\frac{P''_J(1)}{m^6}u^2+...\right\rangle\,,\nonumber\\
k=2:\qquad  g_{2,0} + g_{3,1} u + g_{4,2} u^2 + . . . &=&\left\langle \frac{P_J(1)}{m^4}+2\frac{P'_J(1)}{m^6}u+2\frac{P''_J(1)}{m^8}u^2+...\right\rangle\,,\nonumber\\
k=3:\qquad g_{3,0}+g_{4,1} u+g_{5,2} u^2+... &=&\left\langle \frac{P_J(1)}{m^6}+2\frac{P'_J(1)}{m^8}u+2\frac{P''_J(1)}{m^{10}}u^2+...\right\rangle\,,\nonumber\\
&\vdots &\label{SUsystem}
\eea
with the definition of the high-energy average~\cite{Caron-Huot:2020cmc},
\be
\langle (...) \rangle \equiv\frac{1}{\pi}\sum_J (2J+1)  \int_{M^2}^\infty  \frac{dm^2}{m^2}\rho_J(m^2) (...)\,.
\label{hea}
\ee

Considering  equations with   $k\geq k_{\rm min}$ (that we will take later to be $k_{\rm min}=1,2$), we can relate the IR Wilson coefficients with the UV-averages of derivatives of   $P_J$ in the following way: 
\be
g_{n+l,l}= \frac{2^l}{l!}\left\langle \frac{P_J^{(l)}(1)}{m^{2(n+l)}}\right\rangle\ ,\ \ \  n\geq k_{\rm min}\,\, \text{and}\  \ l=0,1,..., \left[\frac{n-1}{2}\right]\,.
\label{gwilsons}
\ee
Since $P_J^{(l)}(1)\geq 0$, the contributions to \eq{gwilsons}  from the different $J$-states
 are always additive, and therefore $g_{n+l,l}\geq 0$ -- this is a direct consequence of the lack of $s<0$  poles in ${\cal M}(s,u)$. Moreover,   $P_J^{(l)}(1)=0$ for $l>J$ implying that states with $J\leq l$ do not contribute to $g_{n+l,l}$.

In particular,
\bea
g_{n,0}&=&\left\langle \frac{1}{m^{2n}}\right\rangle=\sum_i \frac{g^2_{i\pi\pi}}{m^{2n}_i}
  \ , \ \ \ \ \nonumber\\
g_{n+1,1}&=& \left\langle \frac{{\cal J}^2}{m^{2(n+1)}}\right\rangle=\sum_i \frac{g^2_{i\pi\pi}\, J_i(J_i+1)}{m^{2(n+1)}_i}
\ , \ \ \ \   \nonumber\\
g_{n+2,2}&=& \frac{1}{4}\left\langle \frac{  {\cal J}^4-2{\cal J}^2}{m^{2(n+1)}}\right\rangle
=\sum_i \frac{g^2_{i\pi\pi}\, J_i(J_i-1)(J_i+1)(J_i+2)}{4m^{2(n+1)}_i}
\,,
\label{wilsons}
\eea
where $\mathcal{J}^2 \equiv J(J+1)$.  
Notice that only for $k_{\rm min}=1$ all Wilson coefficients  have a dispersive representation in terms of \eq{gwilsons}. For $k_{\rm min}=2$, the couplings $g_{1,0}$ and $g_{2,1}$ are not captured by these dispersion relations.

\vspace{5mm}

In a similar way,  we can  obtain dispersion relations for the ${\cal M}(s,t)$ amplitude, whose analytic structure is given in Fig.~\ref{fig:tchannel},
\be
\frac{1}{2 i}\oint_{\rm C_0} ds'\ \frac{{\cal M}(s',-u-s')}{s^{\prime k+1}}=
\int_{M^2}^\infty  ds'\ \frac{{\rm Im} {\cal M}(s',-u-s')}{s^{\prime k+1}}
+(-1)^k\int_{M^2}^\infty  ds'\ \frac{{\rm Im} {\cal M}(s',-u-s')}{(s'+u)^{k+1}}.
\label{ST}
\ee
These can be expanded as in the previous section, and provide yet more  relations~\cite{Albert:2022oes}. 
In  the case  $k_{\rm min}>1$  these new relations are crucial, as they give access to Wilson coefficients that do not have a dispersive representation in terms of ${\cal M}(s,u)$.
In particular, for $k_{\rm min}=2$, the coupling $g_{2,1}$ is not determined by \eq{gwilsons} but appears in \eq{ST},
\be
 g_{2,1} =2 g_{2,0}-2 \left\langle \frac{(-1)^J}{m^4}\right\rangle\,,
\label{g21}
\ee
while for $k_{\rm min}=3$, $ g_{3,1}$ can only be determined by
\be
 g_{3,1}=3 g_{3,0} +\left\langle \frac{(-1)^J (2{\cal J}^2-3)}{m^6}\right\rangle\,.
\label{g31}
\ee

\subsection{Null Constraints} 

The dispersion relations  in \eq{SUsystem}, and the small-$u$ expansion of \eq{ST}, over-determine the Wilson coefficients. This  leads to a set of \emph{null constraints},
\begin{equation}
\la {\cal X}_{n,k}(J,m^2)\ra=0\,,\quad \la {\cal Y}_{n,k}(J,m^2)\ra=0\,,
\label{uNC}
\end{equation}
on the high-energy spectral density, with $m^{2n} {\cal X}_{n,k}$ and $ m^{2n}{\cal Y}_{n,k}$ functions of $J$ only. Their  compact expression at all orders is provided in Ref.~\cite{Albert:2022oes}.
%
For the analytic arguments in this article we are only interested in the most relevant null constraints (those involving less powers of $1/m$) and in those with the leading asymptotic $J\to\infty$ behavior at a fixed order $n$ in $1/m^{2n}$.

For $k_{\min}=1$, there is  one (and only one) null constraint $\sim O(J^{2 (n-1)}/m^{2n})$ at each order~$n$,\footnote{We use a slightly different normalization w.r.t. Ref.~\cite{Albert:2022oes}, 
which has no impact on  \eq{uNC}.}
\bea
  n=2:\qquad  m^4\, \mathcal{Y}_{2,1} &=& -2(1-(-1)^J) + \mathcal{J}^2\,,\nonumber\\
  n=3:\qquad  m^6\, \mathcal{X}_{3,1}  &=& -6 \mathcal{J}^2+  \mathcal{J}^4\label{x31}\,,\nonumber\\
  n=4:\qquad  m^8\,  \mathcal{X}_{4,1}  &=& -24 \mathcal{J}^2 -8\mathcal{J}^4+\mathcal{J}^6\,,
  \nonumber\\
  &\vdots &\nonumber\\
   \qquad (n-1)!^2\ m^{2n}\,  \mathcal{X}_{n,1}  &=& 
  \frac{2^{n-1}}{(n-1)!} {P_J^{(n-1)}(1)} -  {\cal J}^2\,.
  \label{nullconst}
\eea
The other null constraints have only subleading  terms in powers of $J$  w.r.t. these.

When we study larger $k_{\rm min}$, the constraints in \eq{nullconst} disappear, and  subleading null constraints  now dominate. For $k_{\rm min}=2$ this involves null constraints that grow as $O(J^{2 (n-2)}/m^{2n})$.
There are two of them at each order $n$, and can be separated into those where the sign of the term  $O(J^{2 (n-2)}/m^{2n})$ is fixed,
and those where this sign oscillates between $J$-odd and $J$-even.
In the first class we have,
\bea
  n=4:\qquad m^8\big(\mathcal{Y}_{4,2}-\mathcal{Y}_{4,1}\big) 
  &=&8(1-(-1)^J)-10\mathcal{J}^2+\mathcal{J}^4\,,\nonumber\\
  n=5:\qquad\qquad\ \ \ \
   m^{10}\, \mathcal{X}_{5,2}  &= &30 \mathcal{J}^2-17\mathcal{J}^4+\mathcal{J}^6\,,\nonumber\\
  n=6:\qquad\qquad\ \ \ \
   m^{12}\, \mathcal{X}_{6,2}  &= &144 \mathcal{J}^2-46\mathcal{J}^4-20\mathcal{J}^6+\mathcal{J}^8\,,\nonumber\\
 \ &\vdots&\nonumber\\
   \qquad  (n-2)!^2\ m^{2n}\,  \mathcal{X}_{n,2}  &=& 
  \frac{2^{n-2}}{(n-2)!} {P_J^{(n-2)}(1)} - 2{P_J^{(2)}(1)}\,.
   \label{nullconstb}
\eea
In both cases, \eq{nullconst} and \eq{nullconstb}, the  $\mathcal{Y}_{n,k}$ null constraints (originating from ${\cal M}(s,t)$ dispersion relations) appear only at the lowest order in $1/m^2$, and at higher order the dominant $J$ behavior is controlled by the  $\mathcal{X}_{n,k}$ null constraints (originating from ${\cal M}(s,u)$ dispersion relations).
On the other hand, the most relevant oscillating null constraint is,
\be
 n=3:\qquad m^6\ \mathcal{Y}_{3,1}\,=\, -6(1-(-1)^J)+2 (1-2(-1)^J) \mathcal{J}^2\,,
 \label{subleading}
 \ee
where  the sign of the $\mathcal{J}^2$ term oscillates  with $J$. 


\begin{figure}[t]
\begin{center}
\includegraphics[width=0.55\linewidth]{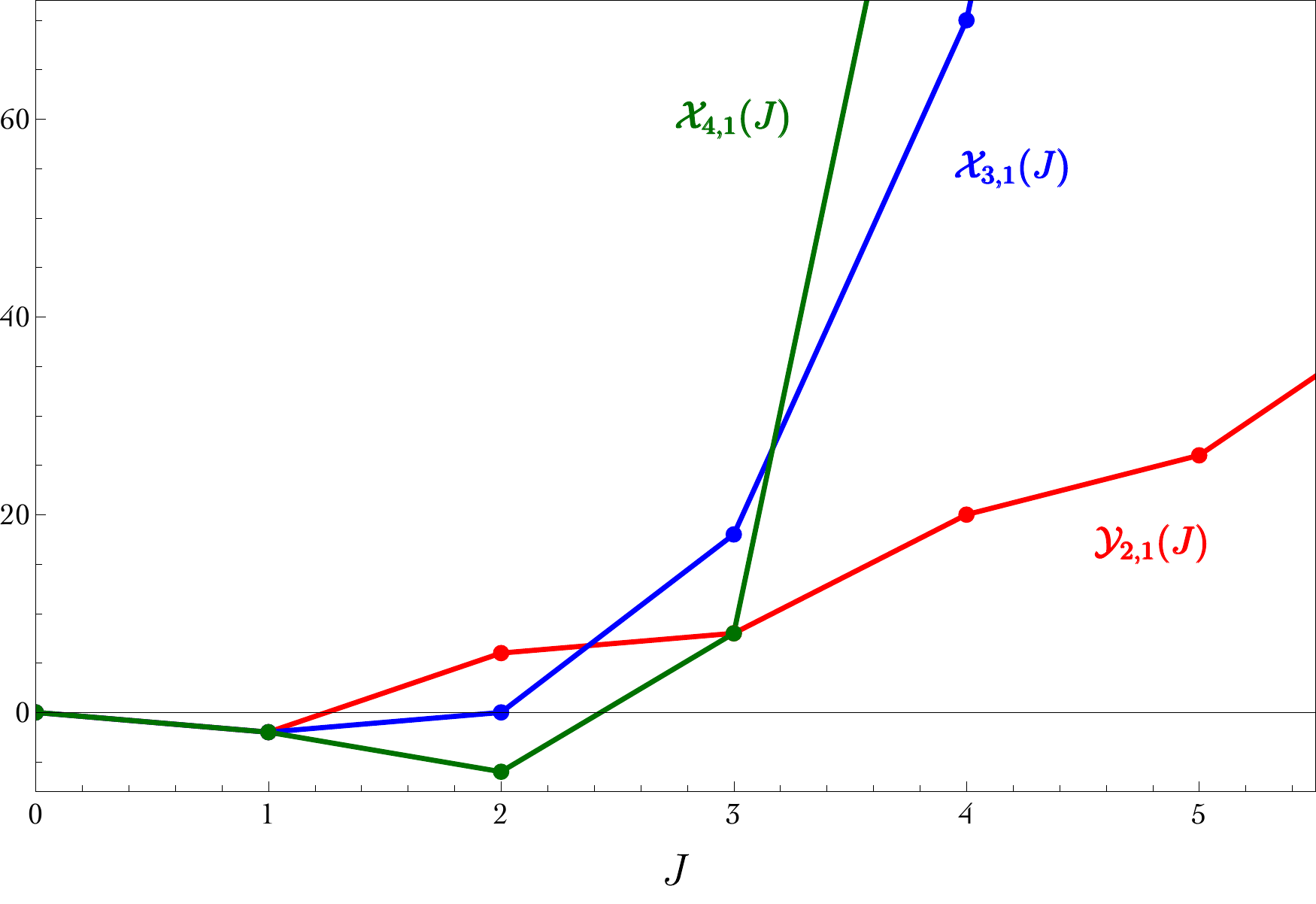}
\caption{\it  Null constraints from   \eq{nullconst},  as a function of $J$, for fixed $m$.}
\label{fig:ncosc}
\end{center}
\end{figure}

Notice that  for $J=0$ the arguments of all null constraints vanish, as 
 can be easily seen in  \eq{nullconst} and \eq{nullconstb} (and more generally by the expressions in Ref.~\cite{Albert:2022oes}).
This implies that  the spin-0 component of the UV spectrum decouples and  is not restricted  by  null constraints.
This  is related  to the fact that models with only $J=0$ states can provide a consistent UV completion
of the pion amplitude, satisfying  \eq{UVa} and \eq{UVb}  for  $k_{\rm min}=1$, as we will discuss in section \ref{sec:UV}.
For $k_{\rm min}=2$,  
also   $J=1$ states give zero contributions and decouple from the null constraints.
This again can be understood from the fact that models with only $J=1$ states  
provide a pion amplitude that consistently satisfies the Froissart-Martin  bound.
This pattern persists:  for $k_{\rm min}= 3$, one finds  that the $J=0,1,2$ states  decouple from 
 the null constraints, and  so on.

It is also instructive to understand how the null constraints can be  satisfied, as this tells us information about the mass spectrum of the theories. 
In Fig.~\ref{fig:ncosc} we show the first few null constraints in \eq{nullconst} as a function of $J$.
We  see that for the first two null constraints the contribution from the $J=1$ is the only opposite in sign
to the other ones. As these expressions must average  to zero, \eq{hea}, this implies that the theory 
must contain $J=1$ states.  For ${\cal X}_{4,1}$, also the  $J=2$ contribution is negative,
implying that  also $J=2$ states are needed.
As the order $n$  of the null constraint increases, one finds that the 
number of states with negative coefficients increases:
eventually   all  $J$ are needed to satisfy the null constraints. 
So, for $k_{\rm min}=1$, theories in the 
large-$N_c$ limit either have no $J>0$ state, or have states with all values of  $J$ from $1$ to~$\infty$. Similarly, for  $k_{\rm min}=2$, states with spin $J\geq 2$ are either absent, or  are all present.

\subsection{UV completions of the pion amplitude}\label{sec:UV}
Before proceeding to examine the implications from positivity, we would like to discuss simple UV completions 
for  a theory of pions.
By this we mean 
theories that generate   consistent crossing-symmetric amplitudes ${\cal M}(s,u)$,  with simple poles at real $s>0$,  positive spectral density,
 satisfying the high-energy behavior of Eqs.~(\ref{UVeq}) for some $k_{\rm min}$.
 Interestingly, these amplitudes  will turn out to  reside  at the  kinks of the allowed parameter space, as we will discuss in the next section.

Simple amplitudes describing exchange of a single spin-$J$ particle with mass $m_J$ are characterised by  a pole-structure  with residue on the associated partial wave,
\begin{equation}\label{amponshell}
{\cal M}^J(s,u)= \frac{m_J^2 P_J(1+2u/m_J^2)}{m_J^2-s}+F(s,u)+(s\leftrightarrow u)\,,
\end{equation}
where $F$ is an analytic function that defines the theory, but  does not contribute to the residue. We will fix $F$ by  imposing the Adler's zero ${\cal M}^{J}(0,0)=0$ and by requiring that the amplitude satisfies the  large-$s$ behavior of Eqs.~(\ref{UVeq}) for the lowest possible value of $k_{\rm min}$.

\subsubsection{Theory of scalars}
\label{scalar}

Scalars can provide a consistent UV completion to  a theory of  pions, via
the  Higgs mechanism in the linear sigma-model.
The $2\rightarrow 2$ pion amplitude mediated by a spin-0 state  with $I=0$ is given by,
\be
{\cal M}^{(s)}(s,u)=\frac{g^2_{s\pi\pi}s}{m^2_{s}-s}+(s\leftrightarrow u)\,.
\label{s}
\ee
Expanding  \eq{s}  at low energies, $s,u\ll m^2_s$, we  obtain,
\be
g_{n,0}=\frac{g^2_{s\pi\pi}}{m^{2n}_s} \ \ \ (n\geq 1)\ ,\quad \quad g_{n,l}=0\ \ \ (l\neq 0)\,.
\label{sw}
\ee
An important property of \eq{s} is that it satisfies the high-energy behavior in Eqs.~(\ref{UVa},\ref{UVb}) with $k_{\rm min}=1$. 
Therefore the Wilson coefficients \eq{sw} obey the sum rules \eq{wilsons}, as can be easily checked.
The fact that a model of only scalars does not need higher-spin states to satisfy Eqs.~(\ref{UVa},\ref{UVb}) with $k_{\rm min}=1$, explains why the $J=0$ states decouple from the null constraints, as explained above.

\subsubsection{Theory of vectors}
\label{vectors}

Let us now consider  a (weakly coupled) spin-1 resonance with isospin $I=1$ (or, in general, in the $Adj.$ representation of $SU(N_f)$), which we will refer to as  $\rho$, in analogy with QCD. 
From \eq{amponshell} we have,
\be
{\cal M}^{(\rho)}(s,u)=\frac{g^2_{\rho\pi\pi}m^2_{\rho}}{m^2_{\rho}-s}P_1\left(1+\frac{2u}{m^2_{\rho}}\right)+(s\leftrightarrow u)\,,
\label{rho}
\ee
corresponding to the contribution  from the transverse components of a massive vector  coupling to pions via $g_{\rho\pi\pi} f_{abc}\rho^{a}_\mu \pi^b\partial^\mu\pi^c$ (minimal coupling) where $f_{abc}$ are the $SU(N_f)$ structure constants.\footnote{A more general function $F$ in \eq{amponshell} would be associated with the exchange of  longitudinal modes (contact terms to the Lagrangian) which would yield in the amplitude multiplicative factors $(s/m_\rho^2)^n$ that worsen the high-energy behavior.
}
This amplitude  can arise in models
in which the $\rho$ gets its mass from the Higgs mechanism, or in holographic models
where the $\rho$ arises as a  Kaluza-Klein state.

\eq{rho}  satisfies the Froissart-Martin bound, Eqs.~(\ref{UVa},\ref{UVb}) but only for $k_{\rm min}=2$.
Nevertheless, the high-energy behaviour of \eq{rho} can be improved  by the following deformation~\cite{Albert:2022oes}:
\be
\widehat{\cal M}^{(\rho)}(s,u)=\frac{g^2_{\rho\pi\pi}m^2_{\rho}}{m^2_{\rho}-s}P_1\left(1+\frac{2u}{m^2_{\rho}}\right)\frac{m^2_\infty}{m^2_\infty-u}+(s\leftrightarrow u)\,,
\label{rhoinft}
\ee
that at high energy $s\gg m^2_\infty$ satisfies Eqs.~(\ref{UVa},\ref{UVb}) with $k_{\rm min}=1$.
By studying the pole structure of \eq{rhoinft}, one can see that the amplitude is mediated by states of any $J$ with masses $m_{\infty}$, but also at  $s=m_{\rho}$ we have now states with $J\not=1$.
Nevertheless, by taking the  limit $m_\infty/m_{\rho}\to \infty$ in \eq{rhoinft},
we  recover  \eq{rho} as well as its low-energy predictions.
 So, as long as we are only interested in the Wilson coefficients, we can safely use  \eq{rho}.

At low energies,
\eq{rho} leads to
\be
g_{1,0}=3\, \frac{g^2_{\rho\pi\pi}}{m^2_\rho} \ ,\ \
g_{2,1}=\frac{4}{3}\,\frac{g_{1,0}}{m^{2}_\rho}\ ,\ \
g_{n,0}=\frac{1}{3}\, \frac{g_{1,0}}{m^{2n}_\rho} \ ,\ \
g_{n+1,1}=\frac{2}{3}\,\frac{g_{1,0}}{m^{2(n+1)}_\rho} \ \ (n\geq 2)\,,
\label{rhocont}
\ee
while $g_{n,l}=0$ for $l\geq2$.   
Alternatively, we could compute the Wilson coefficients \eq{rhocont}  from  dispersion relations, by  using the explicit form of the UV spectral density \eq{weakspectral}, with support on $J=1$ and $m_i=m_\rho$ only. This can help us to appreciate the difference between ${\cal M}^{(\rho)}$ and its improved version $\widehat{\cal M}^{(\rho)}$.
Indeed, since the vector amplitude \eq{rho} fulfils \eq{UVa}  only for $k_{\rm min}\geq 2$, we can not use the sum rules \eq{gwilsons}  for $k_{\rm min}=1$.  
In particular, the expressions for $g_{1,0}$ and  $g_{2,1}$ from \eq{wilsons} do not hold -- indeed they differ from those obtained directly from the amplitude \eq{rhocont}. 
On the other hand, \eq{rho} fulfills \eq{UVb} for $k_{\rm min}=1$ and one can use the prediction for $g_{2,1}$ from  \eq{g21}, that agrees with \eq{rhocont}. 

If instead we use $\widehat{\cal M}^{(\rho)}$ from \eq{rhoinft} -- which has $k_{\rm min}=1$ high-energy behavior -- the extra  states  beyond the $\rho$  give a nonzero contribution to $g_{1,0}$ and  $g_{2,1}$  that makes it to coincide with  \eq{rhocont}.
These contributions tend to zero in  \eq{g21}.

\subsubsection{Theory of spin-2 states}
\label{secf2}

The pole structure of four-pion amplitude   mediated by a spin-2 state only (in analogy with QCD we refer to it as $f_2$), 
can be written as,
\be
{\cal M}^{(f_2)}=\frac{g^2_{f_2\pi\pi}m^2_{f_2}}{m^2_{f_2}-s}P_2\left(1+\frac{2u}{m^2_{f_2}}\right)+(s\leftrightarrow u)\,.
\label{f2}
\ee
Since $P_2(x)=(3 x^2-1)/2$, this amplitude  grows like $\sim s^2/u$ for large $s$,
violating even  the Froissart-Martin $k_{\rm min}=2$ bound. 
Nevertheless, \eq{f2}  can be deformed as in \eq{rhoinft} to improve its high-energy behavior:
\be
\widehat {\cal M}^{(f_2)}=\frac{g^2_{f_2\pi\pi}m^2_{f_2}}{m^2_{f_2}-s}P_2\left(1+\frac{2u}{m^2_{f_2}}\right)\frac{m^2_\infty}{m^2_\infty-u}+(s\leftrightarrow u)\,,
\label{f2infty}
\ee
that indeed satisfies the  Froissart-Martin  bound  for $s\gg  m_\infty^2$, and leads to \eq{f2}
in the   limit $m_\infty/m_{f_2}\to \infty$. {It is not possible to improve this further, as the resulting amplitude would have negative spectral density, and violates unitarity. For this reason amplitudes with light $J\geq 3$ (which grow as $s^J$ from $P_J(1+2s/m_J^2)$ can not be completed into amplitudes that satisfy the Froissart-Martin bound, see also~\cite{Bellazzini:2019bzh}.}

Contact terms  (the function $F$ in \eq{amponshell}) would modify \eq{f2}. 
Nevertheless,  demanding that they do not grow faster than \eq{f2}, leaves only  terms up to $O(s)$. These  terms can only  affect  $g_{1,0}$,  telling us that  this Wilson coefficient cannot be predicted in  a theory of spin-2 states.
The rest of the Wilson coefficients, however, can be unambiguously derived from \eq{f2} in the low-energy limit: 
\bea
&&g_{2,0}=7\, \frac{g^2_{f_2\pi\pi}}{m_{f2}^4}\ ,\ \ \
g_{2,1}=\frac{12}{7}g_{2,0}=g_{3,1}m^2_{f_2}=g_{4,2}m^4_{f_2}\ , \ \ \  
g_{n,0}=\frac{1}{7}\, \frac{g_{2,0}}{m^{2n-4}_{f_2}} \ \ \ (n\geq3)\,,
\nonumber\\
&&
g_{n,1}=\frac{6}{7}\, \frac{g_{2,0}}{m^{2n-4}_{f_2}} \ \ (n\geq4)\ ,\ \ \ 
g_{n,2}=\frac{6}{7}\, \frac{g_{2,0}}{m^{2n-4}_{f_2}} \ \ (n\geq5)\,,
\label{f2cont}
\eea
with $g_{n,l}=0$ for $l>2$. 

Similarly to the spin-1 case, we could calculate some of  the Wilson coefficients from dispersion relations, with  spectral density for $J=2$ only.
As \eq{f2} satisfies the high-energy limit Eqs.~(\ref{UVeq}) for $k\geq k_{\rm min}=3$,  
we can  use the sum rules \eq{gwilsons}  with $k_{\rm min}=3$ to  obtain  all the Wilson coefficients except
 $g_{1,0},\ g_{2,0},\ g_{2,1},\ g_{3,1},\ g_{4,2}$ and $g_{5,3}$ -- indeed their value via
  \eq{gwilsons} disagrees with \eq{f2cont}. 
This can also be understood by realizing that the   states of mass $m_\infty$ present in the consistent amplitude \eq{f2infty}  give  finite contributions to $g_{1,0},\ g_{2,0},\ g_{2,1},\ g_{3,1},\ g_{4,2}$ and $g_{5,3}$ 
when \eq{gwilsons} is used, even in the limit $m_\infty\to \infty$.
For the coefficient $g_{3,1}$, however,  we can  alternatively determine it by  using \eq{g31} that indeed agrees with \eq{f2cont}.



\subsubsection{The $su$-models}

The $su$-models \cite{Caron-Huot:2020cmc,Caron-Huot:2021rmr}  give  the simplest four-pion amplitude mediated entirely by higher-spin states.
The particularity of these models is that their spectrum is fully degenerate,
${\cal M}(s,u)\propto 1/(s-m^2)(u-m^2)$, a condition that naturally places these models at the boundary of the allowed parameter space, as we shall see.
These $su$-amplitudes and the associated Wilson coefficients  are discussed in detail in Appendix~\ref{appc}.

\subsubsection{The Lovelace-Shapiro and Coon amplitude}

There are other four-pion amplitudes mediated by higher-spin states:
 the Lovelace-Shapiro~\cite{Lovelace:1968kjy,Shapiro:1969km} amplitude and its generalization, the Coon amplitude
\cite{Coon:1972qz}.
These amplitudes originate in the context of string theory, and can provide therefore fully consistent UV completions to a theory of pions.  We will see that the Wilson coefficients   
predicted by these amplitudes lie at  the  closest point to one of the boundaries of the allowed regions.
Therefore they  provide information about  the mass spectrum of the theories residing on these boundaries.
The details of these amplitudes are given in Appendices~\ref{appd} and \ref{appe}.

\section{Implications of ${\cal M}(s,u)/s \to 0$ at large $s$}
\label{ms}

In this section, we  assume that the pion  amplitude ${\cal M}(s,u)$  satisfies the conditions Eqs.~(\ref{UVa},\ref{UVb}) for $k_{\rm min}=1$, as argued in Ref.~\cite{Albert:2022oes}. 
Since all Wilson coefficients scale like $1/N_c$, and because positivity bounds are inherently projective 
(i.e. only ratios of Wilson coefficients are constrained), we will work with
\be
\tilde g_{n,l}\equiv \frac{g_{n,l}}{g_{1,0}}M^{2(n-1)}\,,
\label{ratioW}
\ee
where $g_{1,0}$ is the leading Wilson coefficient and $M$ the EFT cutoff defined in Fig~\ref{fig:sumrules}. Unless stated, we will take  $M$ as the lowest resonance mass. 
 The $\tilde g_{n,l}$  are independent of $N_c$ in the large-$N_c$ limit.

We can use the sum rules \eq{gwilsons} to determine the Wilson coefficients as a function of the mass spectrum, which is itself constrained by the null constraints. Our goal will be to shape the boundaries of the EFT parameter space, and  identify  the UV theories that generate it. When possible we shall use analytic arguments, complemented when necessary by numerics.


\subsection{Bounds on the leading Wilson coefficients}
\label{g2s}
\begin{figure}[t]
\begin{center}
\includegraphics[width=0.8\linewidth]{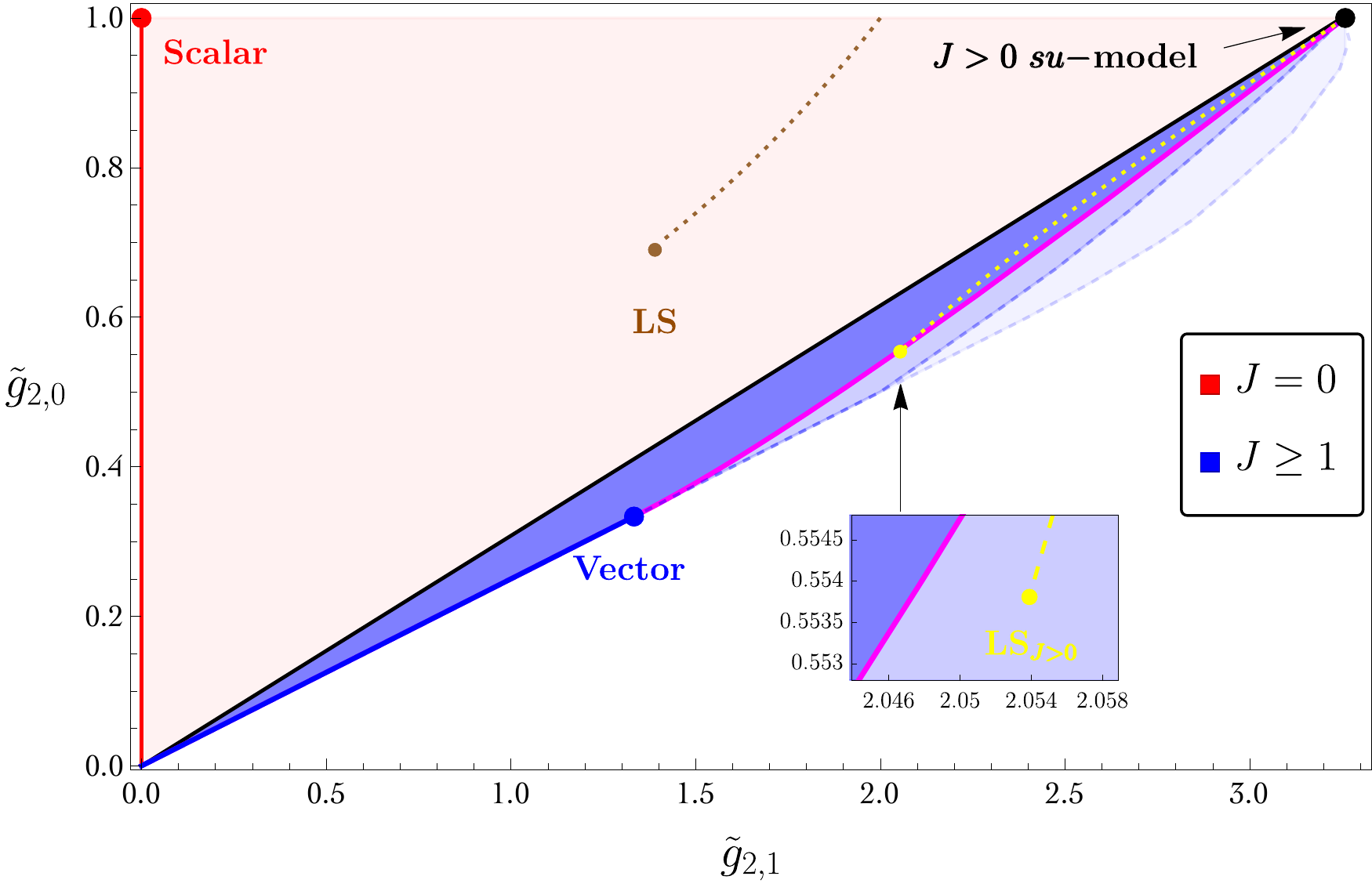}
\caption{\it Allowed regions in the  $\tilde g_{2,1}$--$\, \tilde g_{2,0}$ plane, divided into regions with $J=0$  states only (red line) and $J>0$ states (blue) -- using null constraints  with $\nmax=3$ (dashed light-blue line) and $\nmax=11$ (dashed  blue line). 
The dark blue area spans \eq{interpol} for  $0\leq m_\rho\leq M$ and  $m_\rho\leq m\leq \infty$.  
Moreover, $J=1$ models lie on the blue line,  $J>0$ $su$-models on the black line
(\eq{suamp} with \eq{lambda} and $0\leq m\leq M$), and the  magenta line  corresponds to \eq{sutomrho}.
The  brown  and yellow dot correspond to the  Lovelace-Shapiro amplitude 
with and without scalars (\eq{LS-Samp})) respectively.
The dashed brown and yellow line are the Coon amplitude (\eq{coonamp} with $C=1$)
with and without $J=0$ states respectively.}
\label{fig:g2g2pdec}
\end{center}
\end{figure}

Let us  start by studying the implications of positivity bounds for $\pi\pi$ scattering at order $O(s^2)$, i.e. for the coefficients $g_{2,0}$ and  $g_{2,1}$. 
 As explained in Ref.~\cite{Albert:2022oes},  and recalled in Appendix~\ref{appa}, 
the allowed regions can be obtained by numerical methods, see Fig.~\ref{fig:g2g2pdec}.
These methods work extremely well as explorative tools and give conclusive answers when they rapidly approach known theories, e.g.~\cite{El-Showk:2012cjh}.
 Nevertheless,
they are limited by computer power, and  leave the question open of whether the extrapolation from finite resolution truly reveals a physically meaningful result. Here we will show that for some questions the numerical convergence is too slow, and we proceed by using analytic methods to map as much of the parameter space as possible.

\paragraph{Scalar theories and the $\tilde g_{2,1}>0$ boundary.}
The smallest value of $\tilde g_{2,1}=0$ is saturated by  an amplitude mediated by $J=0$ states,  as discussed in section \ref{scalar} -- see also Ref.~\cite{Albert:2022oes}. In particular,  from \eq{sw}, identifying $M=m_s$,  
we have $(\tilde g_{2,1},\tilde g_{2,0})=(0,1)$, depicted as a red dot in  Fig.~\ref{fig:g2g2pdec}.
This point must   clearly be a corner (kink) of the full allowed region, since  from  \eq{wilsons} one can  see
that $\tilde g_{2,1}\geq 0$  and $\tilde g_{2,0}\leq 1$.
When more (non-degenerate) scalars are present, the value of $g_{2,0}/g_{1,0}$ always decreases, since $g_{2,0}$ scales as $\sim1/m_s^{4}$ and therefore extra heavy scalars  contribute more to $g_{1,0}$
than $g_{2,0}$ (the same is true for $g_{n,0}/g_{n-1,0}$). \label{pagedecrease}
This is shown in  Fig.~\ref{fig:g2g2pdec} by a  red line.

 As discussed in section \ref{sec:disprel}, states with $J<k_{\rm min}$ decouple from the null constraints. For $k_{\rm min}=1$, only the  $J=0$ states decouple, meaning that it is the only simple theory that can provide a standalone UV completion of the chiral Lagrangian. All other UV completions must involve infinitely many states with all spins.
It is therefore interesting to study the boundary of the region with $J\geq 1$ independently, as we discuss in what follows.

\paragraph{Spin-1 theories, the ${\tilde g_{2,1}}/{\tilde g_{2,0}}<4$ boundary  and its kink.}
The largest value of $\tilde g_{2,1}/\tilde{g}_{2,0}$ can be determined
from \eq{wilsons} and \eq{g21}, which gives
\be
\frac{\tilde g_{2,1}}{\tilde g_{2,0}}=\frac{4\left\langle \frac{1}{m^{4}}\right\rangle_{J-\rm odd}}{\left\langle \frac{1}{m^{4}}\right\rangle_{J-\rm odd}+\left\langle \frac{1}{m^{4}}\right\rangle_{J-\rm even}}\,,
\label{ratiog2s}
\ee
where the subscript denotes average over even or odd $J$ only.
This is saturated by  ${\tilde g_{2,1}}/{\tilde g_{2,0}}=4$ and  
corresponds to theories with only $J$-odd states.
The simplest theory  of this type  is that described in Sec.~\ref{vectors}   consisting of a  light spin-1 state only, whose amplitude with  the  improved high-energy behavior  is given in \eq{rhoinft} with $m_\infty/m_\rho\to \infty$. From  \eq{rhocont}  we have 
 \be
 (\tilde g_{2,1},\tilde g_{2,0})_{\rm vector}=(4/3,1/3)\,,
 \label{kink}
 \ee 
 shown in Fig.~\ref{fig:g2g2pdec} as a blue dot.
 Adding extra spin-1 states  allows us to move down from \eq{kink} to the origin,
along the blue line of Fig.~\ref{fig:g2g2pdec}.
This line must be part of the boundary of the allowed region 
for $({\tilde g_{2,1}},{\tilde g_{2,0}})$,  since
the ratio ${\tilde g_{2,1}}/{\tilde g_{2,0}}=4$ takes the largest possible value.

\vspace{3mm}
The important question to address now is whether  the allowed region can extend along the blue line beyond 
\eq{kink} or not.
The numerical analysis of \cite{Albert:2022oes}  was able to show that 
as one increases the number of null constraints, the kink moves towards \eq{kink}, but appeared to tend asymptotically to a larger value.
Here we will show that the kink resides at \eq{kink}, i.e. the extremal theory along the boundary contains  
 only a spin-1 light state. 

To show this, we will first argue that at the boundary,
\be
\tilde{g}_{2,0} =  M^2\frac{ \left\langle \frac{1}{m^4}\right\rangle
}{ \left\langle \frac{1}{m^2}\right\rangle }\quad \to\quad 
M^2\frac{\left \langle \frac{1}{m^4}\right\rangle_{J=1}}{\left\langle \frac{1}{m^2}\right\rangle _{J=1}}\,\frac{1}{ 1+\frac{\left\langle \frac{1}{m^2}\right\rangle_{J>1}}{\left\langle \frac{1}{m^2}\right\rangle_{J=1}}}\,.
\label{eq:g2}
\ee
This can be proven as follows. \eq{ratiog2s} implies that at the  boundary even-$J$ states must decouple from the average in $g_{2,0}$,
\be
\la \frac{1}{m^4}\ra_{J-\rm even}\to 0\,.
\label{first}
\ee
Using the Cauchy-Schwarz inequality (where  $\la {\cal J}^4/m^8\ra$ must be finite because it enters in the null constraint  in the first expression of \eq{nullconstb}), we have
\be
\sqrt{ \left\langle \frac{{\cal J}^4}{m^8}\right\rangle_{J-\rm even} \left\langle \frac{1}{m^4}\right\rangle_{J-\rm even}}\geq\left\langle \frac{{\cal J}^2}{m^6}\right\rangle_{J-\rm even} \to 0\,.
\ee
This, together with the null constraint $\la{\cal Y}_{3,1}\ra=0$ \eq{subleading}, implies,
 \be
\left\langle \frac{{\cal J}^2}{m^{6}}\right\rangle _{J-\rm even}=3\left\langle \frac{{\cal J}^2-2}{m^{6}}\right\rangle_{J-\rm odd}\geq 12\left\langle \frac{1}{m^{6}}\right\rangle_{J>1-\rm odd}\to0\,,
\ee
because ${\cal J}^2-2=0$ on $J=1$ and ${\cal J}^2-2\geq4$ for $J\geq 2$.
Using Cauchy-Schwarz again (with measure on odd $J>1$ only), and the fact that  $\la \frac{1}{m^2}\ra_{J>1-\rm odd}$ is finite, we find
\be
\sqrt{\left\langle \frac{1}{m^6}\right\rangle_{J>1-\rm odd}\left\langle \frac{1}{m^2  }\right\rangle_{J>1-\rm odd}}\geq \left\langle \frac{1}{m^4  }\right\rangle_{J>1-\rm odd} \to 0\,,
\label{Jm6odd}
\ee
thus proving \eq{eq:g2}.
The theory that lives on the boundary must therefore
consist of  spin-1 states at finite mass accompanied  by higher-spin states at  infinite mass.

Unfortunately, this is not yet enough to claim that it is \emph{exactly} the theory in \eq{rhoinft}, since there might be different ways in which the $J\to\infty$ states enter in the spectral density.


To prove that it is, we  reformulate the question of finding the maximum value of the kink, as a 1D moment problem.
Indeed, the kink is positioned at the maximum value of $\tilde g_{2,0}$ along the boundary given by \eq{eq:g2}. The first factor ${\big \langle \frac{1}{m^4}\rangle_{J=1}}/{\langle \frac{1}{m^2}\ra _{J=1}}\leq 1$ is maximally saturated when the spectrum contains only one spin-1  particle at the mass $M$, $M^2\langle \frac{1}{m^4}\rangle_{J=1}=\langle \frac{1}{m^2}\rangle_{J=1}\equiv g_{\rho\pi\pi}^2/M^2$.
 Then, from the second factor in \eq{eq:g2} we read that the kink is located at the \emph{minimum} of,
\begin{equation} \label{mu0}
\frac{M^2}{g_{\rho\pi\pi}^2}\left\langle \frac{1}{m^2}\right\rangle_{J>1} \,.
\end{equation}
This minimum cannot be zero, because the null constraints relate the high-energy averages to $g_{\rho\pi\pi}^2$, in such a way that the ratio is finite. Indeed, along the boundary, the dominant  null constraints \eq{nullconst} can be written as,
\be
2 (n-1)!^2= \frac{M^{2n}}{g_{\rho\pi\pi}^2}\left\langle \frac{J^{2(n-1)}}{ m^{2n}}\right\rangle_{J>1} \ ,\ \ n=\underline{2},3,4,...\,,\\
\label{ncr}
\ee
i.e. all subleading terms  $J^k$ with $k<2(n-1)$ can be neglected. This can be understood from the comment below \eq{Jm6odd} -- for states with infinite mass, if in the average  $\lim_{m\to\infty}J^{2(n-1)}/m^{2n}$  is finite, then all subleading powers of $J$ must vanish.\footnote{It can also be seen by using the H\"older inequality and $\la{1}/{m^{2k}}\ra_{J>1}\to 0$ (for $k\geq 1$) from \eq{Jm6odd},
\be
 \left\langle\frac{1}{m^{2k}}\right\rangle_{J>1}^{\frac{1}{k}} \left\langle \frac{J^{2(n-k-1)}}{m^{2(n-k)}}\right\rangle_{J>1}^{1-\frac{1}{k}}  \geq\left\langle \frac{J^{2(n-k-1)}}{m^{2n}}\right\rangle_{J>1} \to 0\,.
 \ee}
With a change of variables to (the square of) impact parameter $x\equiv  J^{2} M^2/ m^{2}$, and redefining $n$, we can write \eq{ncr} as
\be
 2n!^2=\int^\infty_0 d\mu (x) x^{n}dx\equiv \mu_n\ \ ,\ \ n=1,2,3,...\,,
\label{ncr2}
\ee
with  $d\mu(x)$ a positive distribution. In this language \eq{mu0} is $\mu_0$, and the problem of finding the kink position translates into a 1-dimensional moment problem:
\begin{gather}
\textrm{\it
finding the 
minimum of $\mu_0$ such that $\{\mu_0,\mu_1,\mu_2,\cdots\}$ is a moment series from a} \nonumber\\
 \textrm{\it positive distribution, with $\mu_n=2n!^2$ for $n\geq1$.}
 \label{problem}
\end{gather}

A sufficient condition for this, is that the Hankel matrix $H^{0}_N$, with  $(H^{k}_{N})_{ij}= \mu_{i+j+k}$, for $i,j=0,\dots, \lfloor N/2\rfloor$, 
be asymptotically positive definite~\cite{lasserreJB,Bellazzini:2020cot},
\begin{equation}
\lim_{N\to\infty} H^{0}_N=2\lim_{N\to\infty}\left( \begin{matrix} \mu_0/2 & 1!^2 & 2!^2 & \cdots & n!^2\\ 1!^2 & 2!^2 & 3!^2& \cdots & (n+1)!^2\\ 2!^2 & 3!^2 & 4!^2& \cdots & (n+2)!^2\\ \cdots & \cdots & \cdots & \ddots & \vdots \\ n!^2 & (n+1)!^2 & (n+2)!^2 & \cdots & (2n)!^2\end{matrix} \right)\succ 0\,.
\end{equation}
Equivalently (using Silvester's criterion) this can be rewritten as,
\begin{equation}\label{Hankelratio}
1-\mu_0/2\leq  \lim_{N\to\infty}\frac{\det  \left. H^{0}_{N}\right |_{\mu_0=1} }{\det H^{2}_{N}}\,.
\end{equation}

Explicit evaluation of \eq{Hankelratio} for fixed $N$ enables us to reach smaller and smaller values, going from $\mu_0\approx0.95$ for $N=10$ (equivalent to $\tilde g_{2,0}\approx 0.51$) to $\mu_0\approx1.54$ for $N=700$ (equivalent to $\tilde g_{2,0}\approx0.39$) -- to be compared with $\tilde g_{2,0}\approx0.42$ of Ref.~\cite{Albert:2022oes}.
Computing the asymptotic behavior of determinants of this type is an interesting open problem in mathematics, see e.g.~\cite{BASOR2001214}, motivated by their appearance in random matrix theory  (interestingly, also in relation with QCD and chiral perturbation theory~\cite{Toublan:2000dn,Akemann:2016keq}).
Leaving this for future work, in Appendix~\ref{appb} we take a shortcut and, rather that computing the individual determinants, we focus on the most efficient way of computing the ratio \eq{Hankelratio}, and show that as $n\to\infty$,
\begin{equation}
\mu_0 \rightarrow 2 \quad\quad\textrm{and}\quad\quad (\tilde g_{2,1},\tilde g_{2,0})\to  (4/3,1/3)\,.
\end{equation}
At the kink resides the theory of a single spin-1 state, with the improved high-energy behavior  amplitude
\eq{rhoinft} with $m_\infty\gg m_\rho$.

\paragraph{The su-model and the  boundary for $J\geq 1$ with minimal $\tilde g_{2,1}/\tilde g_{2,0}$.}
At the largest value of $\tilde g_{2,0}=1$ must lie theories with a degenerate spectrum, see \eq{wilsons}. Apart from a theory of a scalar (discussed before),
the only  amplitude with this property is  the $su$-model discussed in Appendix~\ref{appc}, with amplitude \eq{suamp}. 
{This amplitude can also be obtained  analytically by solving the null constraints. Indeed, for a degenerate spectrum, the null constraints reduce to a system of equations for the couplings $g_{i\pi\pi}^2$. The dominant null constraints \eq{nullconst}, for instance, are linearly independent, and can be solved explicitly for a fixed number of couplings $g_{i\pi\pi}^2$ with $i=1,\cdots,n$. The solution is a function that can be resummed and converges into the $su$-model prediction.
}

 This $su$-model contains a fraction of  scalar residues, controlled by the value of $\lambda$ in \eq{lambdarange}; for the value in \eq{lambda} the theory has no scalars. Its amplitude lies at,
\be
 (\tilde g_{2,1},\tilde g_{2,0})_{J>0\ su\rm -model}=(\approx 3.26, 1)\,,
 \label{kink2}
\ee
shown by the black dot in  Fig.~\ref{fig:g2g2pdec}. The uniqueness of this amplitude naturally puts it at kink of the $J\geq 1$ region (and its linear combination with the scalar amplitude at the boundary of the $J\geq 0$ region).

As the spectrum becomes heavier, $M/m\to 0$, the $su$-model morphs into the free theory (at the origin of the plot in  Fig.~\ref{fig:g2g2pdec}).
 Interestingly, this line defines the boundary of the allowed region for theories characterised by resonances of spin $J\geq 1$. 
In Appendix~\ref{appc} we show that this line is indeed a boundary by considering the most generic deformation of the $su$-model and showing that in order to not spoil positivity these deformations must push you in the bulk of the exclusion region (excluding the contribution from scalars).


\paragraph{Boundary between kinks.}
Therefore we are left with the boundary   line joining the two kinks, \eq{kink} and \eq{kink2}.
We have not found an analytical expression for this curve. 
Nevertheless, as proposed in Ref.~\cite{Albert:2022oes}, we can have a reasonable analytical formula by considering a model that interpolates between a spin-1 model and the $J>0$ $su$-model. The interpolating amplitude  is given by\footnote{There are other possible interpolating amplitudes, but \eq{interpol} is the one with the smallest number of states that we have found.  Adding more states will give predictions for $(\tilde g_{2,1},\tilde g_{2,0})$ that will lie on the left of the magenta line  of Fig.~\ref{fig:g2g2pdec}.}
\be
{\cal M}={\cal M}^{(su)}_1-\frac{3(\ln 8-2)}{g^2_{\rho\pi\pi}} \left(\widehat {\cal M}^{(\rho)}(m_\rho\to m)-\widehat {\cal M}^{(\rho)}\right)\,,
\label{interpol}
\ee 
which corresponds to a $J>0$ $su$-model  in which the spin-1 state of mass $m$ has been subtracted and replaced by a  spin-1 state of mass $m_\rho$. 
We have used  the corrected amplitude $\widehat {\cal M}^{(\rho)}$ from \eq{rhoinft}, to assure that  \eq{interpol} satisfies the high-energy conditions with $k_{\rm min}=1$ for all values of its parameters; nevertheless we will be taking the limit $m_\infty/m_\rho \to \infty$
that corresponds to $\widehat{\cal M}^{(\rho)}\to{\cal M}^{(\rho)}$.
From \eq {interpol} it is clear that by varying the mass  $m$ from $m_\rho$ to infinity, we are effectively pushing up the masses of the $J>1$ states in the $su$-model, leaving only a $J=1$ state at low energy.  
The corresponding Wilson coefficients of \eq{interpol} can be easily calculated and one obtains, for~$M=m_\rho$,
\be
\tilde g_{2,0}=\frac{a -(3-10 a)(r^2-1)}{ a -(9-28 a)(r-1)}\ ,\ \ \ \ 
\tilde g_{2,1}=\frac{1+(36a-11)(r^2-1)}{ a-(9-28a)(r-1)}\,,
\label{sutomrho}
\ee
\noindent where $r={m^2_\rho}/{m^2}$ and $a=1-\ln 2$. 
The magenta line  in Fig.~\ref{fig:g2g2pdec} is obtained by varying $r\in [0,1]$.

We have  determined  this boundary numerically (following  Ref.~\cite{Albert:2022oes}, see our Appendix \ref{appa} for details),  including null constraints  with $\nmax=3$ and $\nmax=11$,
  shown by dashed lines in Fig.~\ref{fig:g2g2pdec}.
Due to the lack of computational power, however, we have not been able to  understand how much 
 the true boundary approaches  the analytic boundary \eq{sutomrho} as $n\to \infty$.
Nevertheless, we can  claim that Eq.~(\ref{sutomrho}) cannot coincide with the true bound, as we know of  another consistent   amplitude that lies on the RHS of this line. This is the Lovelace-Shapiro amplitude in which 
the  scalar  contribution has been removed (see Appendix~\ref{appd} for details).  
The prediction for this  amplitude  (see \eq{LSwilsons})
is shown by a yellow dot in   Fig.~\ref{fig:g2g2pdec}.
In spite of this, the prediction of this amplitude is impressively  close to \eq{interpol} 
as can be appreciated by the  zoom  area in   Fig.~\ref{fig:g2g2pdec}.
We also show  a version of the scalar-subtracted Coon amplitude, \eq{coonamp} in Appendix \ref{appe}, with $C=1$ and $q$  varied from $0$ to $1$. This line starts at the $J>0$ $su$-model and goes down to the Lovelace-Shapiro model. It  slightly improves \eq{sutomrho} but only close to the  Lovelace-Shapiro values ($q\simeq 1$).
It could be interesting to know if there is another deformation of the Lovelace-Shapiro amplitude 
that saturates the true boundary.
We also show in Fig.~\ref{fig:g2g2pdec} the prediction from the 
full Lovelace-Shapiro  and Coon amplitudes (with $C=1$) without subtracting the scalar contributions.

For models with both  $J=0$ and $J>0$ states,
the allowed region corresponds to the convex hull spanned by the individual boundaries, shown
 in  light red (plus blue)  in Fig.~\ref{fig:g2g2pdec}.

\subsection{Emergence of Vector Meson Dominance}

We have seen that a theory of  a spin-1 state does not satisfy ${\cal M}(s,u)/s\to 0$ at large $s$,  fixed $u<0$, and requires  higher-spin  states to soften the high-energy behavior. The converse is  also true: any model of higher-spin
states must contain spin-1 mesons. This can be made  explicit by looking at  null constraints. For example, the   first two null constraints,  $\left\langle \mathcal{Y}_{2,1}\right\rangle =0$  and $\left\langle \mathcal{X}_{3,1}\right\rangle =0$ of \eq{nullconst}, lead to 
\bea
\left\langle \frac{1}{m^4}\right\rangle \bigg|_{J=1} &=& 
3\left\langle \frac{1}{m^4}\right\rangle\bigg|_{J=2}+4\left\langle \frac{1}{m^4}\right\rangle\bigg|_{J=3}+\cdots
\nonumber\,,\\
\left\langle \frac{1}{m^6}\right\rangle \bigg|_{J=1} &=& 
9\left\langle \frac{1}{m^6}\right\rangle\bigg|_{J=3}+35\left\langle \frac{1}{m^6}\right\rangle\bigg|_{J=4}+\cdots\,.
\label{sums}
\eea
Since the RHS is always positive, this identity can only be fulfilled if there are spin-1 states in the theory.
These equations also tell us that, at any order in $1/m^n$ in the average, the contributions from any individual $J>1$ state must  always be smaller than the $\rho$ contribution, since the coefficients appearing on the RHS are always bigger than one. 
Moreover, these coefficients scale  with large ${J}$ as  $\sim {J}^{2n-2} /m^{2n}$, which is faster than how they appear in the low-energy couplings such as                                                                                                                                                                                                                                                      $g_{n,0}\sim  \langle 1/m^{2n}\rangle $ or $g_{n,1}\sim \langle J^2/m^{2n}\rangle $ (but $g_{2,1}\sim \langle 1/m^{4}\rangle $) from \eq{gwilsons}.\footnote{For couplings $g_{n,l}$ with larger  and larger $l$ this argument is more involved, since the $J=1$ contribution cancels from \eq{gwilsons} and is restored only via null constraints.}

\begin{figure}[t]
\centering
\hskip-.5cm  \includegraphics[width=0.5\textwidth]{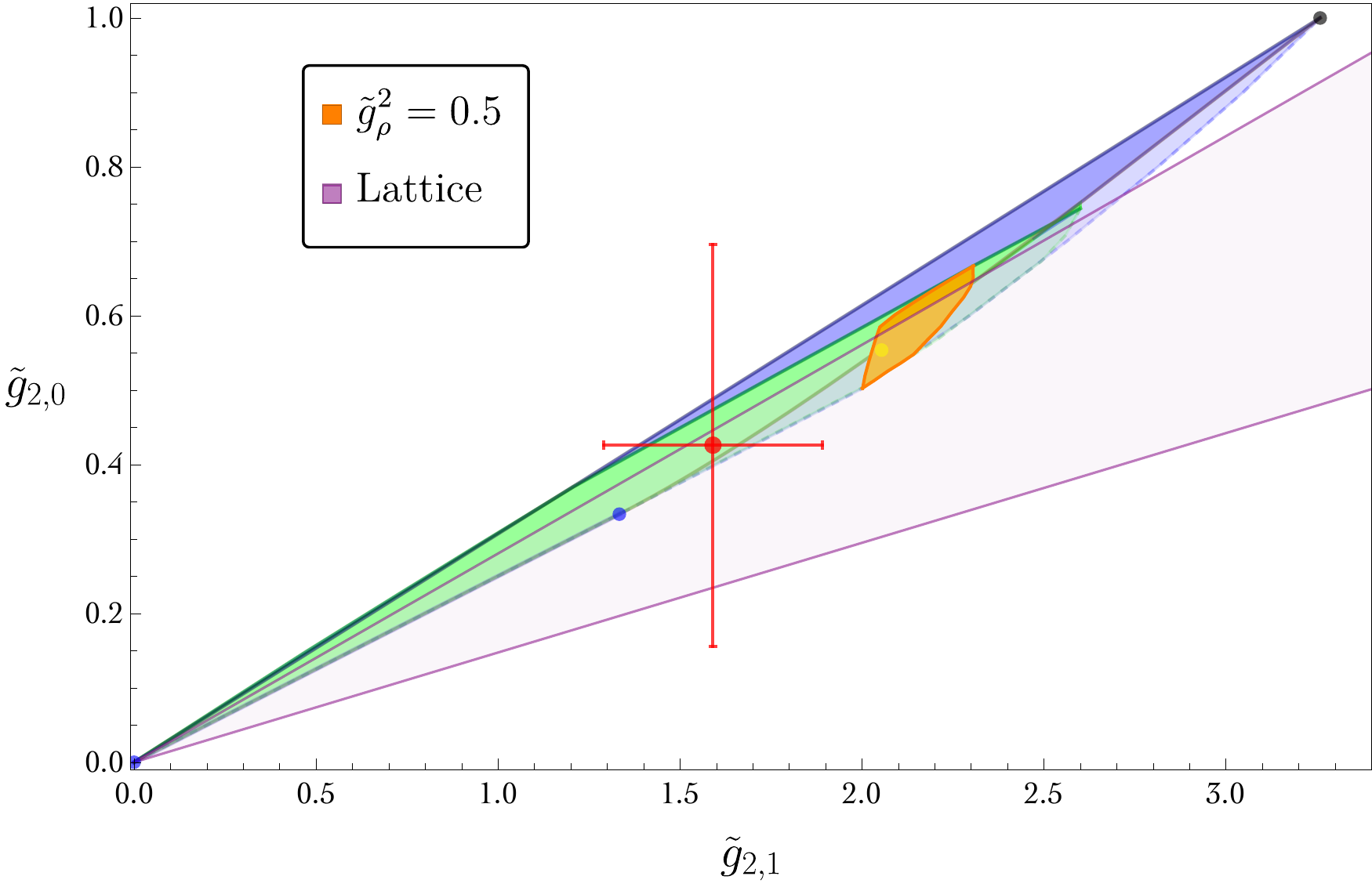}
\includegraphics[width=0.5\textwidth]{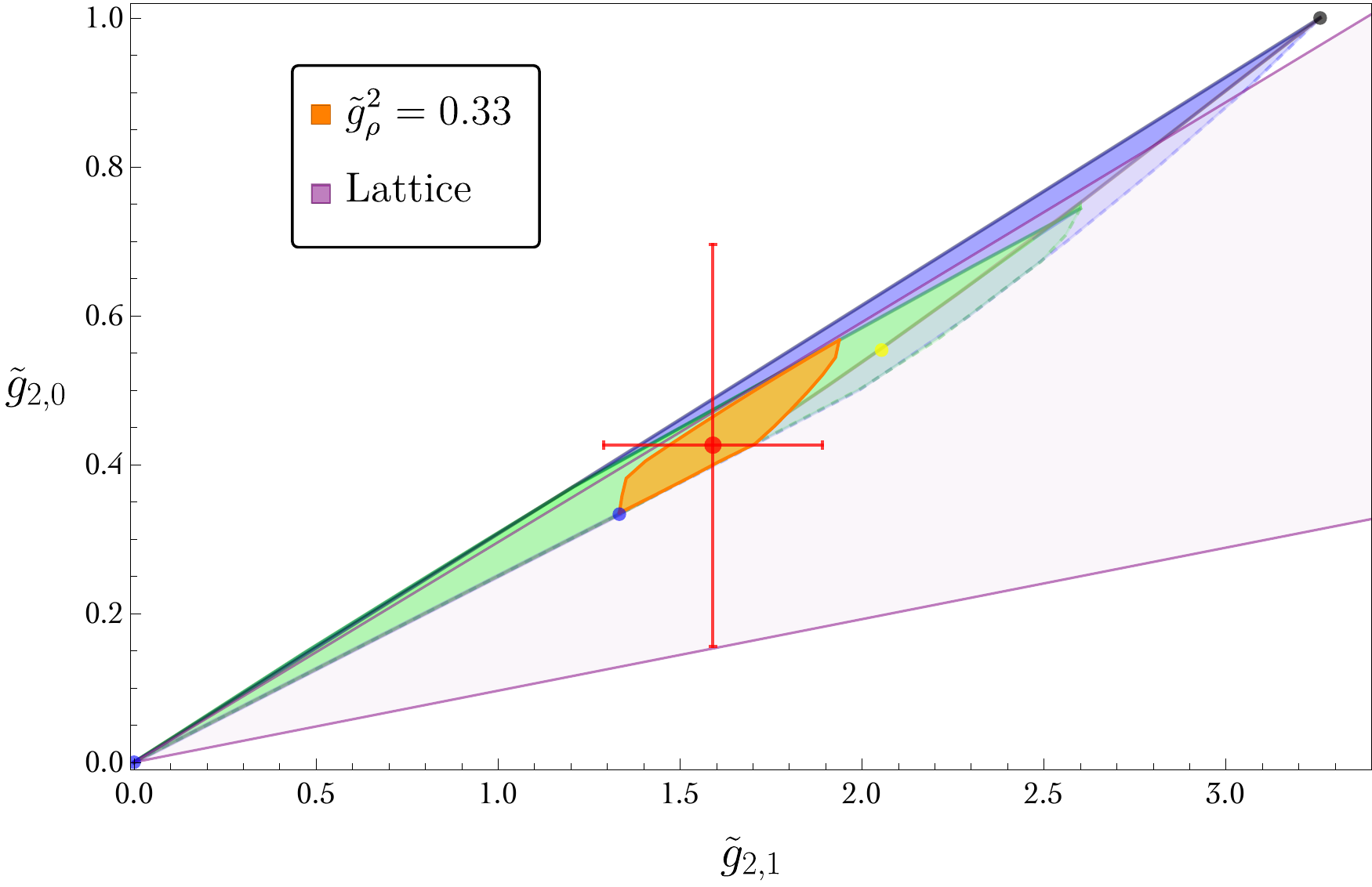}
\caption{\it 
Allowed regions from positivity with $J\geq1$ (as in Fig.~\ref{fig:g2g2pdec}).
The  green regions have heavier  resonances of all spin $J\geq 1$ with masses $M'\geq 1.65\, m_\rho$:  the dashed green line uses null constraints up to $\nmax=11$. 
In orange the allowed region for a fixed $\tilde g_\rho=1/2$ (left) and  $\tilde g_\rho=1/3$ (right), with still $M'/m_{\rho}=1.65$. In purple, the allowed region from Lattice simulations \cite{Bali:2013kia,Hernandez:2019qed,Baeza-Ballesteros:2022azb}. In red the QCD experimental value from Ref.~\cite{Bijnens:1994ie}.}
\label{VMDfixrho}
\end{figure}

The property that the $\rho$ meson dominates the low-energy amplitude of pions (or at least that amplitudes with $\rho$ mesons populate the  space of consistent pion amplitudes) is referred to as Vector Meson Dominance (VMD)~\cite{Sakurai,Ecker:1988te}. Despite its poor  theoretical motivation, VMD is  known to lead to good agreement with  QCD experimental data. 
Here we see that VMD emerges from unitarity and crossing symmetry. For the most relevant couplings, it is illustrated by the alignment  between  the $J>0$ allowed region and the spin-1 line observed   in Fig.~\ref{VMDfixrho}. 
The QCD experimental value, as determined  in Ref.~\cite{Bijnens:1994ie}, is denoted by the red cross in the figure.

As the masses of the 
$J>1$ states  increase, 
VMD becomes more manifest.
Indeed,  
from Eqs.~(\ref{wilsons},\ref{g21}), the contributions from the high-spin  states at large mass  $m_{J>1}\to \infty$ go~as,
 $$\frac{g_{2,0}}{g_{1,0}}\sim \frac{g_{2,1}}{g_{1,0}} \sim \frac{1}{m^2_{J>1}}\to 0\,,$$
 while the $J=1$ contribution remains finite.
This is also true for the rest of the Wilson coefficients, since for large scale separations between the $\rho$ and higher spins  $m_\rho/m_{J>1}\to 0$, The null constraints \eq{nullconst} 
require a finite contribution from the  $J\to \infty$ states, in such a way that  $ J^2/m^2_{J>1}$ remains fixed in units of $ g_{1,0}\sim 1/m^2_{J>1}$.
Using   \eq{wilsons}, this  implies that   the  $J>1$ contribution  to the  Wilson coefficients scales as,
\be
\Delta g_{n+l,l}\sim \frac{J^{2l}}{m^{2(n+l)}_{J>1}} \sim  \frac{1}{m^{2n}_{J>1}} \,,
\ee
and therefore tends to zero for large $m_{J>1}$.


At finite masses, we can study this effect numerically, requiring a finite mass gap between the $\rho$ meson and other
resonances  $M'>m_\rho$, i.e. we work with the spectral density \cite{Albert:2022oes},
 \be
(2J+1)\rho_J(s)\to\frac{\pi}{2}g^2_{\rho\pi\pi}\delta_{J,1}\delta(s-m^2_\rho)+(2J+1) \rho'_J(s)\,,
\label{subrho}
\ee
where the last term corresponds to extra states  with $s\in[M',\infty)$ (see Appendix~\ref{appa}). 
In QCD, the lightest higher-spin resonce is a spin-2 meson $f_2$  
with a mass $m_{f_2}\approx 1.3$ GeV, which implies  $M'/m_{\rho}\sim 1.65$.
 With this mass gap, the allowed region reduces to  the small green strip  in Fig.~\ref{VMDfixrho}, which is more strongly aligned with the  the spin-1 prediction, evidencing VMD.
 

 It is instructive to further divide the allowed region of parameters in terms of the contribution of the $\rho$ to the leading Wilson coefficient $g_{1,0}$, i.e.
\be
\tilde{g}^2_\rho\equiv \frac{g^2_{\rho\pi\pi}}{g_{1,0}m_\rho^2}\,,
\ee
which can be thought to quantify VMD.
 In Fig.~\ref{VMDfixrho} we show in orange the allowed regions in which $\tilde{g}^2_\rho$  matches the experimental QCD value and
the value taken in the vector model \eq{rho},
\be
 (a)\ \tilde{g}_\rho^2 \simeq  \frac{1}{2}  \ \ \ (\text{QCD})\ ,\ \ \ \
 (b)\  \tilde{g}_\rho^2 = \frac{1}{3}\ \ \ (\text{spin-1 model})\,.
\label{fixingrho}
 \ee
As we will discuss in Sec.~\ref{coups}, $\tilde{g}_\rho^2 =  {1}/{2}$ is close to its maximal value $\tilde{g}^2_\rho\sim 0.78$, which is also where the associated spectral density has the $J>1$ contributions  maximised (saturated by the $su$-model).
On the other hand, even 
for $\tilde{g}_\rho^2 =  {1}/{3}$ (for which $2/3$ of the leading effects are taken care by higher spin-mesons), the allowed region for $\tilde g_{2,0}$ and $\tilde g_{2,1}$ still sits close to the spin-1 contribution (blue dot),
showing  a small effect from the $J>1$ states.

Our discussion of VMD so far focused on quantifying the contributions from $J>1$ mesons: these are the difficult ones to model, and for which our arguments are particularly important.
 On the other hand, since $J=0$  states decouple from the null constraint, they  could indeed dominate 
 $\tilde g_{2,0}$ and $\tilde g_{2,1}$, as it happens in the Higgs model.
Nevertheless,  scalars can be  easily accommodated in any phenomenological analyses as they
 have simple UV completions.
It is worth noticing, however, 
 that when a spin-1 $\rho$ is assumed to be the lightest  meson in the spectrum,  as in QCD,
the scalar contribution becomes  smaller.
This property is tied to the fact that  contributions  to the Wilson coefficients are always positive. 
For example,  taking scalars with masses $\gtrsim 1.65\ m_\rho$, 
while still fixing $\tilde{g}_\rho^2$ to the values considered above, we find  that the 
resulting allowed regions depicted in Fig.~\ref{VMDfixrho}
increase in size by only $10-25\%$ along the $\tilde g_{2,0}$ direction.


\subsubsection{Comparison with Lattice QCD}
\label{QCDcon}

 The Wilson coefficients $L_{1,2,3}$, traditionally defined in the $SU(3)$ chiral Lagrangian~\cite{Gasser:1984gg},\footnote{Following the chiral Lagrangian definition in Ref.~\cite{Gasser:1984gg},
\be
{\cal L}=\frac{F_\pi^2}{4}\text{Tr}\left(\partial_\mu U^\dagger \partial^\mu U\right)+L_1\text{Tr}^2\left(\partial_\mu U^\dagger \partial^\mu U\right) + L_2 \text{Tr}\left(\partial_\mu U^\dagger \partial_\nu U\right)\text{Tr}\left(\partial^\mu U^\dagger \partial^\nu U\right)  +L_3 \text{Tr}\left(\partial_\mu U^\dagger \partial^\mu U \partial_\nu U^\dagger \partial^\nu U\right)\,.
\nonumber
\ee}
 are related to ours by
 \be
\tilde g_{2,0}=4(2L_1+3 L_2+L_3)\frac{M^2}{F_\pi^2}\ , \ \quad \tilde g_{2,1}=16L_2\frac{M^2}{F_\pi^2}\,.
\ee
In the large-$N_c$ limit,  $2L_1=L_2$ \cite{Gasser:1984gg}, which leads to
	\be
	\frac{\tilde g_{2,0}}{\tilde g_{2,1}}=\frac{1}{4}\left(1+\frac{\Delta_L}{L_2}\right) ,
	\label{ratio}
	\ee
where $\Delta_L=3L_2+L_3$. This quantity vanishes for theories with only spin-1 resonances, so VMD predicts $\Delta_L\sim 0$. 
Moreover, the positivity of the Wilson coefficients implies that $L_2,\Delta_L\geq 0$. 

A combination of recent lattice simulation results for
large-$N_c$ QCD  \cite{Bali:2013kia,Hernandez:2019qed,Baeza-Ballesteros:2022azb}, gives
  $\Delta_L=(-0.12\pm 0.22)\cdot10^{-3}$.
Unfortunately, we have not found any lattice determination of $L_2$. 
Nevertheless, for any given  value
of $\tilde g_\rho$, we have a minimal value for $L_2\propto g_{2,1}$, which 
plugged into \eq{ratio} with the lattice value of $\Delta_L$, 
can provide   a  bound on ${\tilde g_{2,0}}/{\tilde g_{2,1}}$.   
 For the two values of \eq{fixingrho},  we find 
  \be
 (a)\  0.15 \lesssim\frac{\tilde g_{2,0}}{\tilde g_{2,1}}\lesssim 0.28\ ,\ \ \
 (b)\  0.10 \lesssim\frac{\tilde g_{2,0}}{\tilde g_{2,1}}\lesssim 0.30\,,
 \ee
that correspond to the purple areas in Fig.~\ref{VMDfixrho}.
We notice that in both cases positivity bounds are complementary to bounds from lattice, ruling out different regions. In the particular case $(a)$, the upper bound coming from lattice seems to be more restrictive  than the one from positivity, but this is not the case as we decrease $\tilde g_\rho^2$.
In the future a combination of both approaches can lead to a better determination of the geometry of the allowed regions in parameter space.

\subsubsection{Holography}
The results in Figs.~\ref{VMDfixrho} provide also an explanation for the success of holography for predicting QCD properties \cite{Erlich:2005qh,DaRold:2005mxj}. Holographic models consist in weakly-coupled 5D constructions describing 4D strongly-coupled dynamics in both the $N_c\to \infty$ limit, and  the limit of a large mass gap between spin-0,1 states and other higher-spin states.\footnote{The model also has 5D gravitons, but these correspond to  glueballs of spin $J\leq 2$, which decouple from the pion amplitudes.}

In real QCD, however,  the mass ratio between the $q\bar q$ mesons of spin-2 and spin-1 is not large, $m_{f_2}/m_{\rho}\sim 1.65$. Therefore, one would expect   holographic models  not to provide a good description of low-energy  QCD, contrary to what is observed \cite{Erlich:2005qh,DaRold:2005mxj}. 
Nevertheless, the above analysis shows  that unitarity, causality and crossing-symmetry suppress
 the effects of higher-spin states in the QCD Wilson coefficients. 
 Therefore, even if the mass gap  $m_{f_2}/m_{\rho}$ is not large, the low-energy QCD quantities are mostly affected by only spin-0 and spin-1 states, which are the  ones captured by holographic models. For this reason they can provide a good fit to real-world QCD. 
 
 In particular, in the holographic model of Ref.~\cite{Panico:2007qd}, one can show that the predictions for $\tilde g_{2,0}$ and $\tilde g_{2,1}$ are very close to those of the vector model
  $(\tilde g_{2,1},\tilde g_{2,0})=(1.32,0.33)$.

\subsection{Higher order Wilson coefficients}
The features that sculpt  the allowed region of  $\tilde g_{2,0}$ and $\tilde g_{2,1}$ play a dominant role also in understanding   higher-order Wilson coefficients, $\tilde g_{n,0}$ versus $\,\tilde g_{n,1}$. From \eq{wilsons} we have,
\be
\frac{\tilde g_{n,1}}{\tilde g_{n,0}}=
 \frac{\la \frac{{\cal J}^2}{m^{2n}}\ra}{\la \frac{1}{m^{2n}}\ra}\,,
  \label{gns}
\ee
whose minimal value (zero) corresponds to a model with $J=0$. 
Focusing instead on $J>0$ theories, the  minimal value arises for  models of spin-1 that give
${\tilde g_{n,1}}/{\tilde g_{n,0}}=2$.\footnote{Notice that we could not use this argument for the case $n=2$, since we cannot use the sum rules in \eq{gns} with $n=2$
for a theory of  $J=1$ states only, as explained  at the end of Sec.~\ref{vectors}. In other words, the infinitely heavy $J>1$  states give zero contribution to \eq{gns} only when $n>2$.}

\begin{figure}[t]
\centering
\includegraphics[width=.8\textwidth]{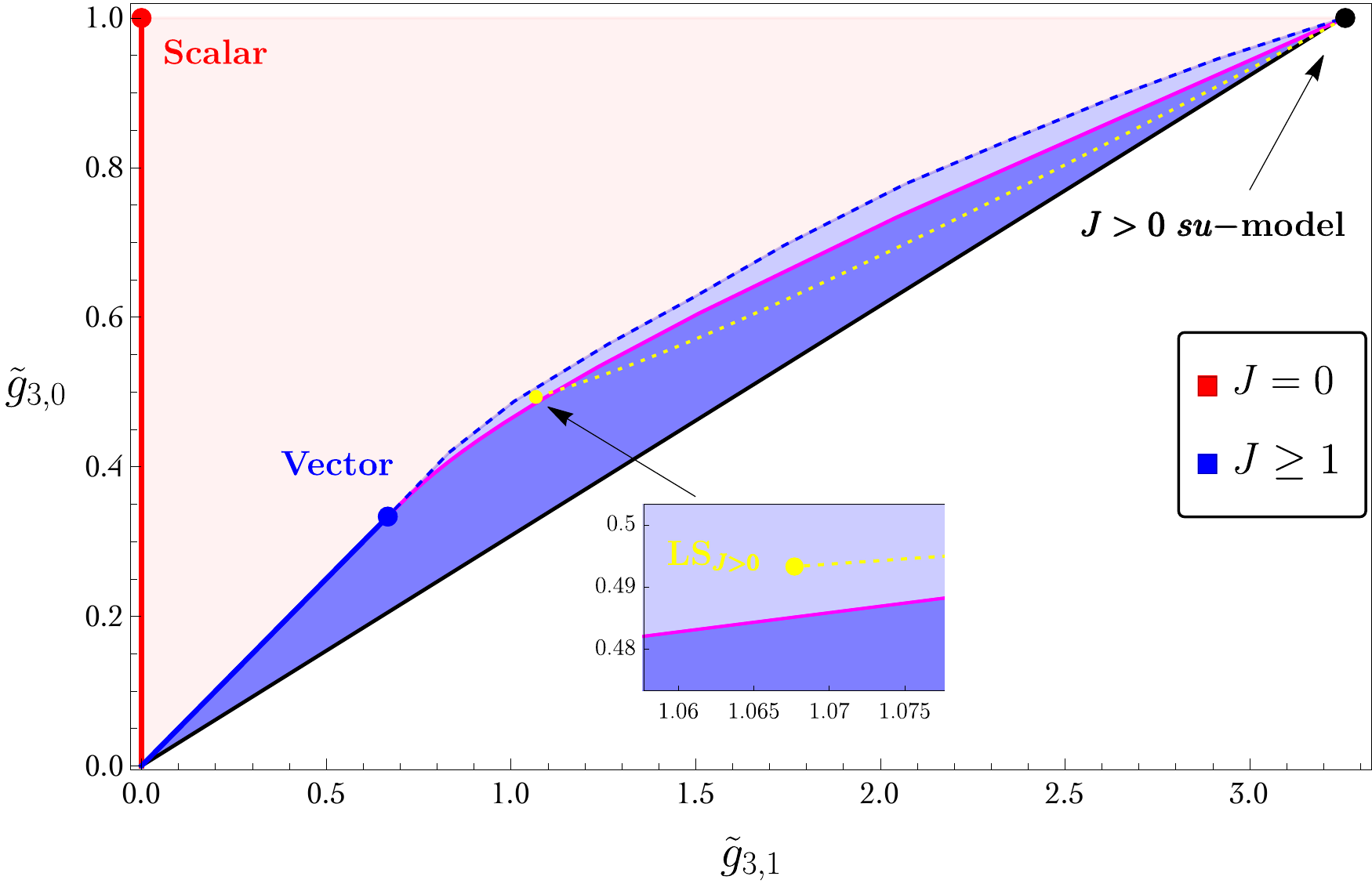}
\caption{\it Allowed region in the $\tilde g_{3,1}$--$\, \tilde g_{3,0}$ plane from positivity. Same  labelling as in  Fig.~\ref{fig:g2g2pdec}, with $\nmax=11$ null constraints.}
\label{g3splot}
\end{figure}

We illustrate this in Fig.~\ref{g3splot}, where a blue dot corresponds to a model with a single  $J=1$ state:
 \be
 (\tilde g_{3,1},\tilde g_{3,0})_{\rm vector}=(2/3,1/3)\,,
 \label{kink3}
 \ee 
 while theories with many spin-1 states populate the blue line.
 
As for $\tilde g_{2,0},\tilde g_{2,1}$,
 we can  show that \eq{kink3}  corresponds to a kink of the boundary, see Appendix~\ref{appb}. 
 The other kink  corresponds again to the $J>0$ $su$-model (the only one with a degenerate spectrum) that gives
 \be
 (\tilde g_{3,1},\tilde g_{3,0})_{su\rm -model}=(\approx 3.26, 1)\,.
 \label{kink4}
 \ee 
We have not been able to find an analytic formula for the boundary connecting the two kinks, \eq{kink3} and \eq{kink4}; we illustrate the numerical analysis in Fig.~\ref{g3splot}. 
We believe  that by adding more null constraints the  boundary  
must approach, but not reach,  \eq{interpol}, consisting of  a theory  connecting the two kinks (the  magenta line in Fig.~\ref{g3splot}).
Nevertheless, as in Sec.~\ref{g2s}, this line cannot be the true boundary
since  the Lovelace-Sphapiro model with $J>0$ states lies at the left of this line, and so does part of the Coon amplitude, \eq{coonamp} (with $C=1$ and $q\in [0,1]$, after subtracting all scalars).



\subsection{Bounding the couplings of mesons to pions}
\label{coups}

\begin{figure}[t]
\centering
\includegraphics[width=0.7\textwidth]{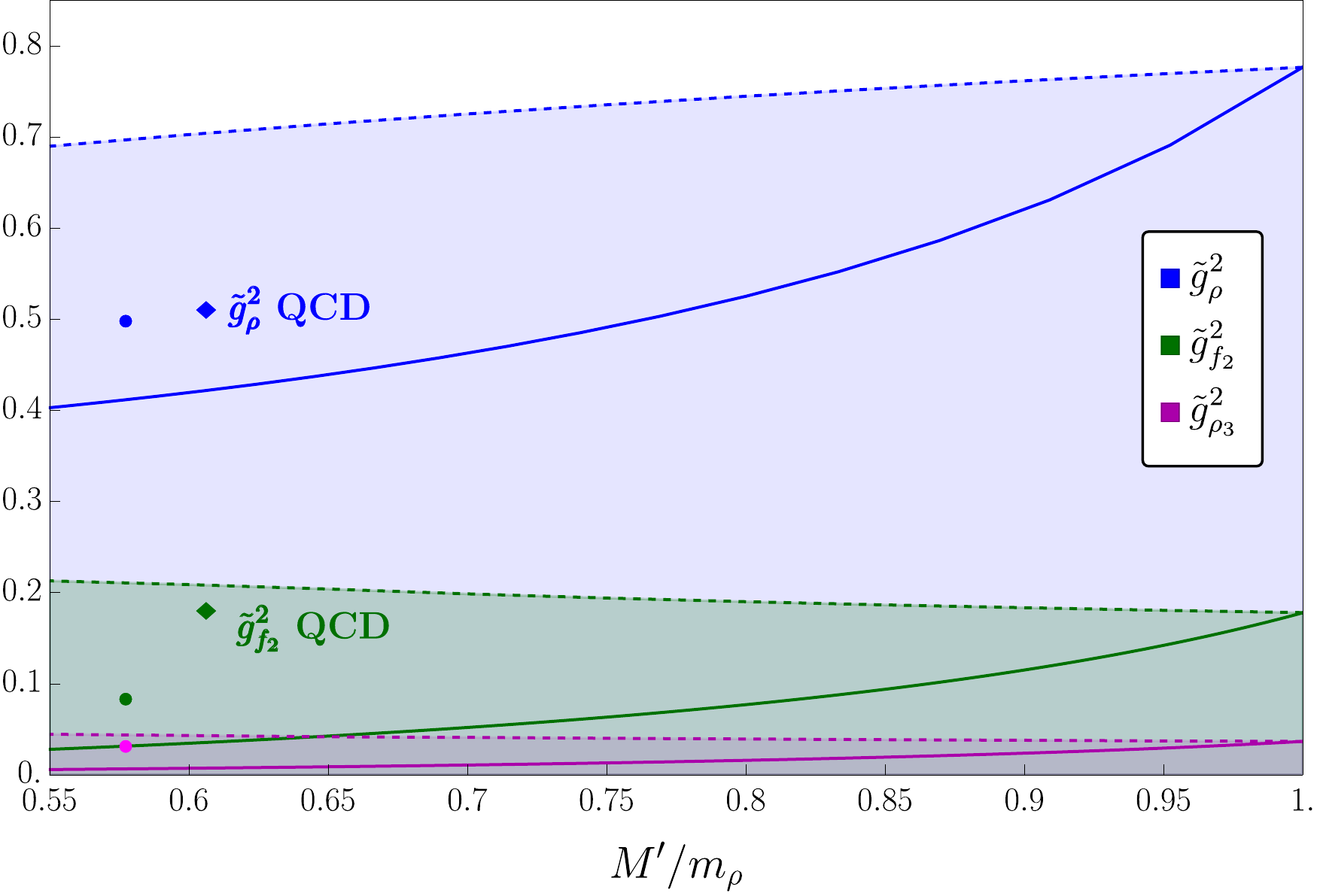}
\caption{\it Upper bound  on   $\tilde{g}_\rho^2$ (dashed blue line), $\tilde{g}^2_{f_2}$  (dashed green line) 
and $\tilde{g}^2_{\rho_3}$ (dashed magenta line) as a function of $M'/m_\rho$ using null constraints with
 $\nmax=7$.
 The solid lines correspond to the prediction from the interpolating model \eq{interpol}. 
The   dots correspond to the values of the Lovelace-Shapiro amplitude without scalars, and 
the diamonds  to the QCD experimental values.
}
\label{fig:couplings}
\end{figure}

So far, we have phrased dispersion relations as UV$\to$ IR vehicles to reformulate microscopic unitarity, causality and crossing-symmetry as predictions for low-energy coefficients. Null constraints, however, provide genuine UV-UV relations, inspired by the same principles. As such, they contain information on the UV meson spectrum and couplings to pions.
We  define the latter, normalized as,
\be
\tilde{g}^2_{i}=\frac{g^2_{i\pi\pi}}{g_{1,0}\, m_i^2}\,,
\label{gi}
\ee
where $i=s,\rho,f_2,\rho_3,...$ labels   $J=0,1,2,3,...$ mesons, following the QCD notation of Ref.~\cite{ParticleDataGroup:2022pth}.

Since spin-0 mesons  decouple from the null constraints, it is easy to understand that $\tilde g^2_{s}$  is maximised by  the smallest possible value of $g_{1,0}$ that, due to its additive property, occurs when the spectrum contains one scalar only:
\be
\tilde g^2_{s}\leq 1\,.
\ee

On the other hand,   bounds on the couplings of $J\geq1$  mesons, involve all null constraints, which we explore  numerically, as 
explained in  Appendix~\ref{appa}. 
The results are illustrated in   Fig.~\ref{fig:couplings}.
For instance, the bound on the $\rho$ coupling $\tilde{g}_\rho^2$ (dashed blue line), is obtained as a function of $M'\geq m_\rho$ by singling out this state from the spectral density as in \eq{subrho}.
The bound goes from a maximal value $\tilde g^2_\rho\simeq 0.78$, 
corresponding to the $J>0$ $su$-model where $M'=m_\rho$,
to the minimal value corresponding to the vector model, $\tilde g^2_\rho=1/3$ where $M'\to \infty$.
This can be compared with the interpolating model \eq{interpol} shown by the solid blue line and  with the  Lovelace-Shapiro model without scalars  
\eq{LS-Samp} (shown by dots) that lies between the two lines.
Interestingly, while all these models give similar predictions in terms of the Wilson coefficients -- see Fig. \ref{fig:g2g2pdec} --  they differ substantially
at the quantitative level in Fig.~\ref{fig:couplings}. 
This provides an interesting  experimental handle to differentiate these theories by testing the couplings of pions to the accessible resonances (amplitude's residues).

Similarly for the spin-2 meson $f_2$, we  rewrite  the spectral density as,
\be
(2J+1)\rho_J(s)\to\frac{\pi}{2}g^2_{\rho\pi\pi}\delta_{J,1}\delta(s-m^2_\rho)+\frac{\pi}{2}g^2_{f_2\pi\pi}\delta_{J,2}\delta(s-M'^2)+(2J+1) \rho'_J(s)\,,
\label{eq:f2}
\ee
and  look for the upper bound on $\tilde{g}_{f_2}^2$ as a function of $M'/m_\rho$ for any
value of  $g^2_{\rho\pi\pi}>0$. This is shown by   the green dashed line in Fig.~\ref{fig:couplings}. 
The resulting bound  is much stricter than that for the $\rho$ -- another manifestation of VMD.

As we study mesons of higher and higher spin $J$, the bound  on $\tilde g^2_i$ becomes stronger and stronger -- see the magenta dashed line for $\tilde{g}_{\rho_3}^2$.
Interestingly, this pattern is also tracked by the QCD experimental values,
$\tilde{g}_\rho^2 = 0.51 \pm 0.01$ and $\tilde{g}_{f_2}^2 = 0.18 \pm 0.01$,
extracted from the widths $\rho,f_2\to \pi\pi$ \cite{ParticleDataGroup:2022pth}, as  shown by the diamonds in Fig.~\ref{fig:couplings}.

\section{Implications of ${\cal M}(s,u)/s^2 \to 0$ at large $s$}
\label{Ms2}

A more conservative assumption is  that the four-pion amplitude satisfies the Froissart-Martin bound, Eqs.~(\ref{UVa},\ref{UVb}) for $k_{\rm min}=2$, rather than $k_{\rm min}=1$ as in the previous section.
 Unfortunately in this case we lose the sum rule for the Wilson coefficient  $g_{1,0}$ (and $g_{2,1}$ is no longer expressible by  \eq{gwilsons} -- which comes from $k_{\rm min}=1$ -- but requires \eq{g21}). 
So, we will need to choose another Wilson coefficient to normalize the rest and make the predictions $N_c$-independent.
 We will use  $g_{2,0}$ and define,
\be
\bar g_{n,l}\equiv \frac{g_{n,l}}{g_{2,0}}M^{2(n-2)}\,.
\label{ratioW2}
\ee 

There is another  important difference w.r.t. Sec.~\ref{ms}. The amplitude mediated by a spin-1 state, \eq{rho}, fulfils ${\cal M}^{(\rho)}/s^2 \to 0$ for large $|s|$, and therefore now provides a good UV description of the  four-pion amplitude. As a consequence, models of spin-1 states do not require anymore  higher-spin states. From the  perspective of null constraints this is realised by the decoupling of $J=1$, when setting $k_{\rm min}=2$, as we explained in Sec.~\ref{sec:disprel}

It is the spin-2 state that now plays  an analog role to that of the $\rho$-meson in Sec.~\ref{ms}. 
Indeed, the $\langle\mathcal{Y}_{4,2}-\mathcal{Y}_{4,1}\rangle=0$ null constraint
reads,
\be
 6\left\langle \frac{1}{m^8}\right\rangle \bigg|_{J=2} = 
10\left\langle \frac{1}{m^8}\right\rangle\bigg|_{J=3}+50\left\langle \frac{1}{m^8}\right\rangle\bigg|_{J=4}+\cdots,
\label{spin2dom}
\ee
which tells us that  spin-2 states need $J>2$ states and viceversa (for instance, the amplitude \eq{f2} requires $J\geq3$ states as in \eq{f2infty}, to comply with the  Froissart-Martin bound).

\subsection{Bounds on Wilson coefficients}

\begin{figure}[t]
\centering
\hskip-.4cm 
  \includegraphics[width=.5\linewidth]{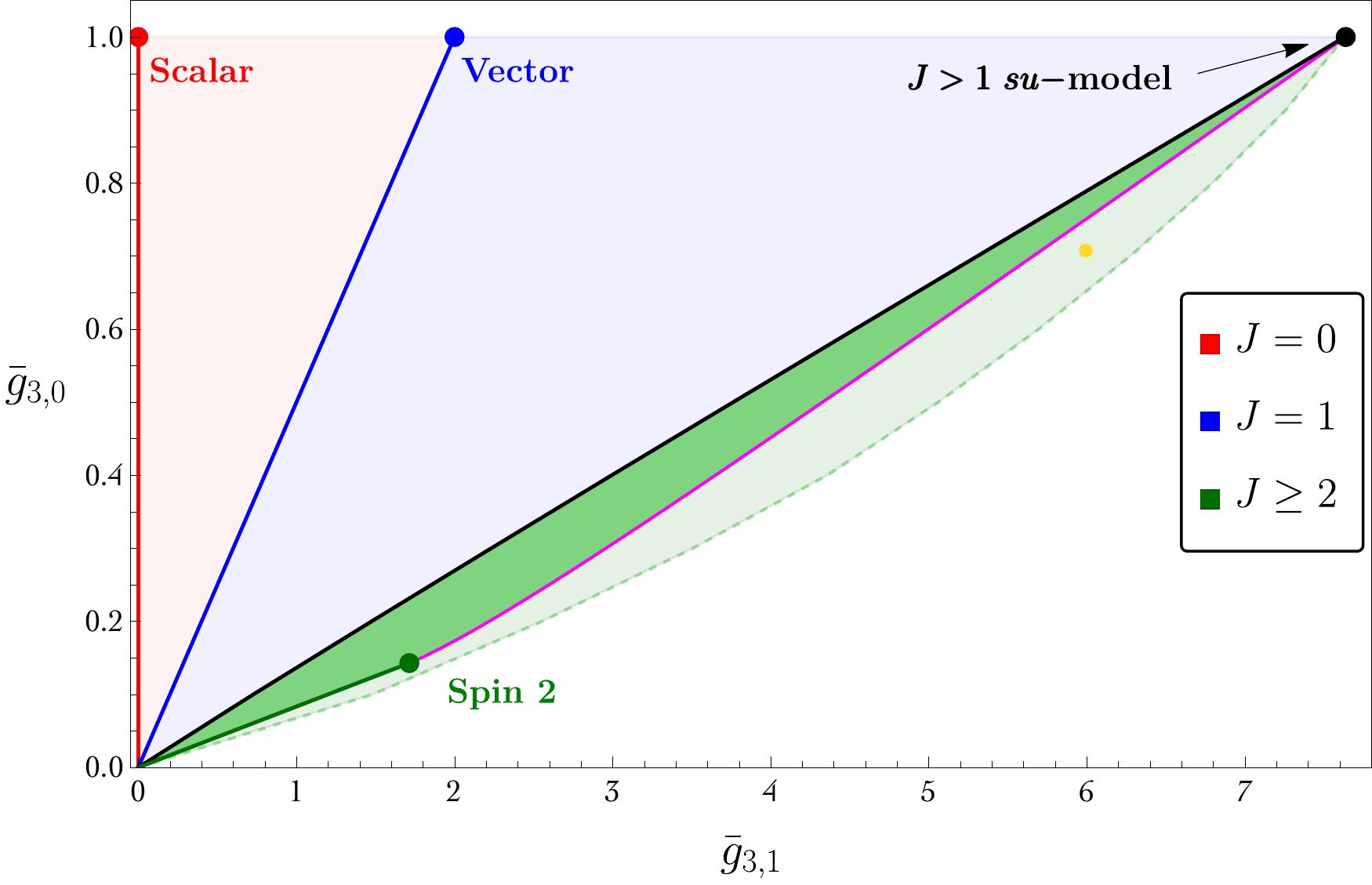}
  \includegraphics[width=.5\linewidth]{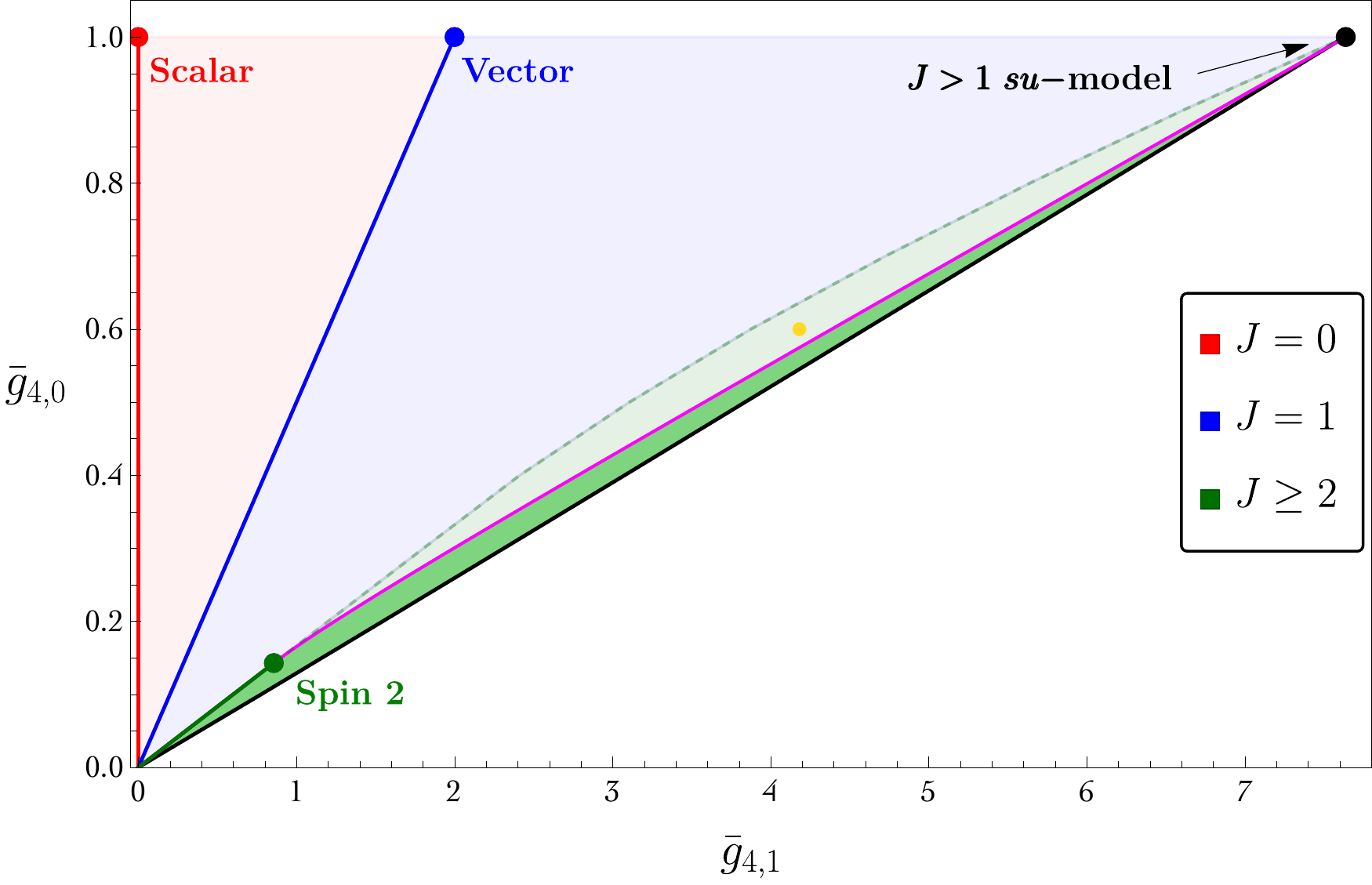}
\caption{\it 
Allowed region in the $\bar g_{3,1}$--$\,\bar g_{3,0}$ plane (left) 
and $\bar g_{4,1}$--$\,\bar g_{4,0}$ plane (right) from positivity. 
The black lines correspond to the $J>1$ $su$-model
(\eq{suamp2} with \eq{lambdas} and  $0\leq m\leq M$),
 the   green lines  to $J=2$ models, and the magenta lines to \eq{sutomf2g3}  and \eq{sutomf2g4} for the left and right plot respectively.
The green areas limited by the dashed line corresponds to the allowed region with $\nmax=7$ null constraints.
}
\label{g34s}
\end{figure}

 At the leading order $O(s^2)$, we only have $\bar g_{2,1}$. 
 As discussed before,  we have $\bar g_{2,1}=0$  ($\bar g_{2,1}=4$) for $J=0$ ($J=1$) models,
that are decoupled from higher-spin states.
 Focusing instead on theories with $J\geq2$ states, the largest value of $\bar g_{2,1}$ comes from the $su$-model, once we have
 subtracted not only the scalar  but also the $J=1$ state,  whose amplitude is given by \eq{suamp2} in appendix~\ref{appc}, using \eq{lambdas}.
 Then, from \eq{su-s-v} we find,
 \be
  \bar g_{2,1}\leq \frac{18\ln 2-13}{10\ln 2-7}\simeq 7.6\,.
\ee  

At order $O(s^3)$ and $O(s^4)$ we can also consider
 $(\bar g_{3,1}, \bar g_{3,0})$ and $(\bar g_{4,1},\bar g_{4,0})$.
The contributions from models of  scalars and vectors are  
given by the lines going respectively from the points $(0,1)$  and $(2,1)$   to the origin,
illustrated in Fig.~\ref{g34s} in red and blue.

The allowed regions   for $J\geq 2$ are less trivial, and correspond to the green areas in Fig.~\ref{g34s}.
Again the upper kink is associated with  the $su$-model, with its degenerate spectrum that
makes $\bar g_{n,0}$ maximal. The coefficients are
 \be
  (\bar g_{n,1},\bar g_{n,0})_{J>1\ su\rm -model}=(\approx 7.6, 1)\,,
 \ee
as illustrated by the black dots in Fig.~\ref{g34s}.
For $m\in[0,M]$, the $su$-model spans the minimum and maximum of the $J\geq 2$ region of  $(\bar g_{3,1}, \bar g_{3,0})$ and $(\bar g_{4,1},\bar g_{4,0})$ respectively (black lines in the figures).

From \eq{gns}   the minimal values for $\bar g_{n,1}/\bar g_{n,0}$ correspond to models with the lowest possible spin. 
This would mean a theory with $J=2$ mesons only, but we have seen that this is incompatible with the Froissart-Martin bound: 
$J\geq 2$ states are also needed to satisfy the null constraints. 
Although these states  can be made infinitely heavy (see section \ref{secf2}), 
 they only decouple in $\bar g_{n,1}$ for $n\geq 4$.
Therefore the $J=2$ theory would give the minimal value for 
 $\bar g_{4,1}/\bar g_{4,0}=J(J+1)|_{J=2}=6$, but not for  $\bar g_{3,1}/\bar g_{3,0}$.
The amplitude of a single spin-2 meson gives from \eq{f2},
 \be
 (\bar g_{4,1},\bar g_{4,0})_{\rm spin-2}=(6/7,1/7)\,.
 \label{kink5}
 \ee 
Following a similar analysis as in Appendix~\ref{appb}, we expect this to be also a kink in the limit where all null constraints are
taken into account. 
The boundary between this kink \eq{kink5} and the $J>1$ $su$-model has be obtained numerically and is illustrated by the green dashed line in~Fig.~\ref{g34s}.

 For  $(\bar g_{3,1},\bar g_{3,0})$, we cannot  assure the existence of  a kink at the $J=2$ value  $(12/7,1/7)$, 
but the numerical bound  (green dashed line) seems to approach this point.

A  consistent amplitude that interpolates 
between the $J>1$ $su$-model and the spin-2 model is given by
\be
\widehat{\cal M}={\cal M}^{(su)}_2-\frac{
5(13\ln2 - 9)}{g^2_{f_2\pi\pi}} \left(\widehat {\cal M}^{(f_2)}(m_{f_2}\to m)-\widehat {\cal M}^{(f_2)}\right)\,,
\label{interpol2}
\ee 
and corresponds to the $J>1$ $su$-model  with  the spin-2 state of mass $m$ removed
and added  back at mass $m_{f_2}$. In the limit $m_\infty\to \infty$, \eq{interpol2} leads  to
\bea
\bar g_{3,0}&=&\frac{20-23r^3+a(-65+75r^3)}{140-143r^2+5a(-91+93r^2)}\ ,\ \ \ \ 
\bar g_{3,1}=\frac{240-245r^3+a(-780+798r^3)}{140-143r^2+5a(-91+93r^2)}\,,\label{sutomf2g3}\\
\bar g_{4,0}&=&\frac{20-23r^4+a(-65+75r^4)}{140-143r^2+5a(-91+93r^2)}\ ,\ \ \ \ 
\bar g_{4,1}=\frac{120-125r^4+6a(-65+68r^4)}{140-143r^2+5a(-91+93r^2)}\,.
\label{sutomf2g4}
\eea
where $r={m^2_{f_2}}/{m^2}$ and $a=1-\ln 2$,
shown in Fig.~\ref{g34s}  as  magenta lines, and compared to  the predictions from the Lovelace-Shapiro amplitude (yellow dot), that   lies -- again -- outside of the area  limited by the magenta lines.

\subsection{VMD and Spin-2 dominance}

As we have seen,  for amplitudes satisfying ${\cal M}/s^2\to 0$ at large energies,  spin-1 states decouple and our previous arguments for VMD no longer hold. In this case,  spin-1, spin-0 and higher spin contributions are independent  phenomenological quantities and VMD would be a mere  accident of Nature.
In spite of this, we still expect  the $f_2$ coupling to pions to be larger than that for $J\geq3$ states, in analogy to VMD. 
To quantify this statement, we define
\be
\bar{g}^2_{i}=\frac{g^2_{i\pi\pi}}{g_{2,0}\, m_i^4}\,,
\label{gi2}
\ee
and  look for  their largest allowed values compatible with positivity. For scalars and vectors, that  decouple from the null constraints, we have
\be
\bar{g}^2_{s},\bar{g}^2_{\rho}\leq1\,.
\ee
For the spin-2 state, instead, we obtain
\be\label{maxg22}
\bar{g}_{f_2}^2\lesssim 0.80\,.
\ee
Similarly to the spin-1 meson in the case $\M/s^2\rightarrow 0$, this latter bound is saturated by the $J>1$ $su$-model.
The QCD experimental value (extracted from data in \cite{ParticleDataGroup:2022pth}) is  $\bar{g}_{f_2}^2 = 0.41 \pm 0.25$ 
and is  much smaller than \eq{maxg22}. 
This is due to the presence of the $\rho$-meson that also  contributes to the normalization factor $g_{2,0}$
and therefore makes $\bar{g}_{f_2}^2$ smaller.
For higher-spin states, we find, taking $m_{i}/m_{f_2}\sim 1.3$,
\be
\bar{g}_{\rho_3}^2\lesssim 0.14\ ,\quad  \bar{g}_{f_4}^2\lesssim 0.04\,.
 \ee

 Interestingly, if we assume $\bar g_{\rho}\neq 0 $ (which is not implied by the null constraints), we can still find a bound on the maximal contribution 
to the Wilson coefficients arising  from $J>1$ states:
\be
\frac{\bar g_{2,1}}{\bar g_{2,1}|_{\rho}}\lesssim 1+ 0.17\, \frac{1-\bar g^2_\rho}{\bar g^2_\rho}\,.
\label{VMD2}
\ee 
Here we have used that  theories containing  $J\geq2$ mesons have $\langle 1/m^4\rangle_{J-\rm odd}/\langle 1/m^4\rangle\lesssim 0.17$, with the bound  saturated by the $J>1$ $su$-model.
Taking for instance $\bar g^2_\rho\sim 0.5$, we obtain from \eq{VMD2} that  the $\rho$ contributes $\gtrsim 80\%$: it still dominates, in line with VMD.

 \section{Conclusions}

We have studied pion scattering amplitudes in the large-$N_c$ limit, using dispersion relations based on crossing-symmetry,  unitarity and causality of the QCD dynamics.
Under different assumptions about the high-energy behavior of amplitudes, but agnostic of the specific meson spectral density, we have identified the allowed regions of parameter space for the  low-energy Wilson coefficients.

For amplitudes with ${\cal M}(s)\lesssim s$ at high-energy, building upon Ref.~\cite{Albert:2022oes}, we have made progress in several directions. By separating theories with  $J=0$ mesons only, from theories with $J\geq 1$, we have revealed a pattern of kinks that characterizes {both the leading and the more irrelevant Wilson coefficients.} Some kinks are populated by known theories, such as the linear sigma model (involving only a spin-0 meson in the UV completion) and the (scalarless) $su$-model, that involves a degenerate spectrum of infinitely many higher-spin resonances.
The other kinks  appear to be associated with a more complex high-energy spectrum, but a numerical exploration of the constraints converges slowly. In these points, we have been able to solve the null constraints analytically, by translating them into a 1D moment problem, and shown that at the kink sits a theory with a single light spin-1 resonance, the $\rho$ meson.

There is one part of the boundary of the allowed region of Wilson coefficients that still eludes analytic methods: here we have found that the scalar-subtracted Lovelace-Shapiro (and part of the  Coon)  amplitude,  provides the best analytical   approximation of the constraints -- within the numerical bounds but outside the region spanned by simpler theories.
This suggests that possibly another deformation of the Lovelace-Shapiro amplitude exists which better approximates the entire boundary.

We have also shown how the soft high-energy behavior of amplitudes implies, via positivity bounds, Vector Meson Dominance and  explains the success of holographic QCD.
The reason is that although these models do not incorporate higher-spin states, positivity constraints
tell us that their contributions in low-energy quantities have to be small.

For amplitudes that are not as soft at high-energy, but that still respect the Froissart-Martin bound ${\cal M}(s)\lesssim s \log^2s$, we have found that analogous results hold.  Now, theories with states with $J=0$ or $J=1$ decouple from the rest and can be studied in isolation. The remaining theories with $J\geq2$ also have kinks, populated by  the (scalarless and vectorless) $su$-model  or by a theory with a single light spin-2 meson. In this context, we have highlighted a version of ``spin-2 meson dominance''.

The comparison between our results and large-$N_c$ lattice simulations (see Fig.~\ref{VMDfixrho}) emphasizes the complementarity between the two methods and lies down an exciting symbiotic program that combines analytic and lattice methods to corner QCD. 

The physics captured by the chiral Lagrangian is  rich and describes many processes beyond $\pi\pi$ scattering: the program  of  cornering large-$N_c$ QCD with positivity bounds
can move forward in many directions.
An important  direction for further  study would be processes involving  pions interacting
with external sources, scalars or photons.
 Here we expect that including full unitarity of the QCD amplitude, as done by Ref.~\cite{Karateev:2022jdb} in the context of the $a$-anomaly, would reveal an interesting interplay between  the chiral anomaly, the QED minimal coupling, and the parameters studied in this article.
This direction of research is interesting also in the framework of physics beyond the Standard Model, in particular for the question of whether the Higgs is composite or not. There, Ref.~\cite{Giudice:2007fh} proposed a set of power-counting rules (on of which was dubbed \emph{minimal coupling}) which are   realized at weak coupling and in string theory -- it would be interesting to prove this in the wider context of dispersion relations. 

On a different front, it would be interesting to incorporate more systematically real-world QCD data to shape  the meson spectrum, or vice-versa, see e.g.~\cite{Caprini:2011ky}. To this goal, it is essential to first  understand finite $N_c$ effects, as discussed in the single-flavor case in Refs.~\cite{Bellazzini:2020cot,Bellazzini:2021oaj} -- a task that we will leave for future work~\cite{Lombardo:2022mel,falkoInprep}.

From a more technical point of view, it is important to bring the  analytic methods we have used in this work into a more systematic tool to identify all theories at the boundary, e.g. via a clever use of Lagrange multipliers and resummations at all orders. This would allow us to understand if the Lovelace-Shapiro amplitude (with subtracted scalars, or scalars and vectors) is indeed extremal or not. Here it would be useful to  further understand which deformations of known amplitudes are consistent with unitarity, along the lines of~\cite{Cheung:2022mkw,Geiser:2022exp,Huang:2022mdb}.

\section*{Acknowledgments}
{F.R. thanks  Jan Albert, Adam Falkowski, Felipe Figueroa, Davide Lombardo, Marcos Marino, Leonardo Rastelli, Antonio Rodriguez, Jullian Sonner, Matteo Tacchi and Piotr Tourkine for important discussions.}
F.S. thanks Filippo Nardi for useful conversations.
F.S. and F.R. acknowledge the support of the European Consortium for Astroparticle Theory in the form of an Exchange Travel Grant.
We are grateful to the Mainz Institute for Theoretical Physics (MITP) of the Cluster of Excellence PRISMA+ (Project ID 39083149), for its hospitality and support during the workshop \emph{Amplitudes meet BSM}.
C.F. is supported by the fellowship FPU18/04733
from the Spanish Ministry of Science, Innovation and Universities.
A.P. has been supported by the Catalan ICREA Academia Program, and the grants 2014-SGR-1450, PID2020-115845GB-I00/AEI/10.13039/501100011033.
The work of F.R.  is supported by the Swiss National Science
Foundation under grants no. 200021-205016 and PP00P2-206149.

\appendix
\section{Numerical Bootstrap}
\label{appa}
In this appendix, we briefly explain the numerical optimization procedure used in this article, following \cite{Caron-Huot:2020cmc,Caron-Huot:2021rmr,Albert:2022oes}. 
From \eq{uNC} we know that,
\be
\la \mathcal{X}_{n,l}(m^2,J) \ra =0 \ , \quad \la \mathcal{Y}_{n,l}(m^2,J) \ra =0\ ,
\ee
while \eq{gwilsons} implies,
\be
g_{n+l,l}=\la g_{n+l,l}(m^2,J) \ra\ , \qquad \text{where} \qquad g_{n+l,l}(m^2,J)= \frac{2^l}{l!} \frac{P_J^{(l)}(1)}{m^{2(n+l)}}\ .
\ee
If we define the vectors 
\be
\vec{v}_1=\begin{pmatrix}
\ 1\ \\ 
0\\
\vdots\\
0
\end{pmatrix}\ ,\quad
\vec{v}_n=\begin{pmatrix}
\ 0\ \\ 
1\\
\vdots\\
0
\end{pmatrix} \ , \quad
\vec{v}_\text{HE}(m^2,J)=\begin{pmatrix}
\ -g_{1,0}(m^2,J) M^2\ \\ 
-g_{n,l}(m^2,J) M^{2n}\\
\mathcal{Y}_{2,1}(m^2,J)\\
\vdots
\end{pmatrix},
\ee
then the following  equation holds
\be
g_{1,0}M^2 \vec{v}_1+g_{n,l}M^{2n} \vec{v}_n+ \la \vec{v}_\text{HE}(m^2,J)\ra =0\ .
\label{bootstrapeq}
\ee
This equation can be adjusted to include more Wilson coefficients.
Now,  multiplying everything by a vector $\vec{\alpha}$ we can solve the following optimization problem:
\bea
&\text{maximize}& \qquad \qquad \vec{\alpha}\cdot \vec{v}_1\nonumber\\
&\text{such that}&\qquad \qquad \vec{\alpha}\cdot\vec{v}_\text{HE}(m^2,J) \geq 0\\
&&\qquad \qquad \vec{\alpha}\cdot\vec{v}_n = \pm 1\ ,\nonumber
\label{genoptprob}
\eea
where positivity of the high-energy average and the normalization $\vec{\alpha}\cdot\vec{v}_n=+1$ will yield the upper bound $\tilde{g}_{n,l}\leq - \vec{\alpha}_{(+)}\cdot \vec{v}_1 $, while $\vec{\alpha}\cdot\vec{v}_n=-1$  the lower bound $ \tilde{g}_{n,l} \geq \ \vec{\alpha}_{(-)}\cdot \vec{v}_1$.

A problem of this kind is solved using a semidefinite problem solver such as \texttt{SDPB} \cite{Simmons-Duffin:2015qma}.
Notice that in this procedure, we divided everything by the positive term $g_{1,0}M^2$ and for the last term in \eq{bootstrapeq} we reabsorbed it in the definition of the high-energy average. Furthermore \texttt{SDPB} can only impose the positivity condition in \eq{genoptprob} at the level of polynomials, therefore we must perform the change of variables $m^2\rightarrow M^2(1+x)$ with $x\geq 0$ and absorb the common denominator once again in the definition of the high-energy average. We can also implement a mass gap simply by modifying the above change of variables to the one $m^2\rightarrow M'^2 (1+x)$, where $M'/M$ is the mass gap. 
Numerically, \eq{bootstrapeq} is evaluated on a grid in $x$ and $J$. The choice of the maximum value $J_\text{max}$  depends on the number of null constraints we want to consider; as more null constraints are included, a bigger $J_\text{max}$ must be chosen in order to ensure the convergence of the bounds. 

In the specific case of constraining $\tilde{g}_{2,0},\tilde{g}_{2,1}$, we input fixed values of $\tilde{g}_{2,0}$, and use
\be
\vec{v}_1+\tilde{g}_{2,0} \vec{v}_2+ \la \vec{v}_\text{HE}(m^2,J)\ra =0\ .
\ee
Then we probe the allowed values of $\tilde{g}_{2,0}$ to find  bounds on $\tilde{g}_{2,1}$,
\be
 (1+\tilde{g}_{2,0})\vec{v}_1+\tilde{g}_{2,1} \vec{v}_2+ \la \vec{v}_\text{HE}(m^2,J)\ra =0\ .
\ee
To extract a $\rho$ state  from the high-energy average and impose a mass gap between the $\rho$ and the spectrum in the UV (as explained in the paragraph above), we define $\vec{v}_\rho=\vec{v}_\text{HE}(m_\rho^2,1)$. Then we start by bounding $\tilde{g}_{2,0}$ via,
\be
\vec{v}_1+\tilde{g}_\rho^2 \vec{v}_\rho+\tilde{g}_{2,0} \vec{v}_2+ \la \vec{v}_\text{HE}(m^2,J)'\ra =0\ .
\ee
We can both reabsorb the term $\tilde{g}_\rho^2 \vec{v}_\rho$ in $\vec{v}_1$ if we want to assume a value for the coupling, otherwise we can add it to the high-energy average in the positivity condition \eq{genoptprob}. Then we use the allowed values of $\tilde{g}_{2,0}$ to bound $\tilde{g}_{2,1}$ with the equation 
\be
 (1+\tilde{g}_{2,0})\vec{v}_1+\tilde{g}_\rho^2 \vec{v}_\rho+\tilde{g}_{2,1} \vec{v}_2+ \la \vec{v}_\text{HE}(m^2,J)'\ra =0\ .
\ee
The procedure described above gives the plots shown in Figs.~\ref{fig:g2g2pdec},\ref{VMDfixrho}. As we vary $n_\text{max}$ (the number of null constraints included), features of the exclusion plots vary.

\section{Minimum of $\mu_0$}
\label{appb}

\emph{A} measure that reproduces \eq{ncr2} is $d\tilde\mu(x)\equiv 4 K_0(2\sqrt{x})dx$, where $K_0$ is the Bessel function of the second kind.
The (Stieltjes) moment problem associated with \eq{problem} is indeterminate, in the sense that there exist multiple measures $d\mu(x)$ that give rise to this sequence of moments. For instance, asymptotically $K_0(\sqrt{x})\to~e^{-\sqrt{x}}$ has the same moments as $e^{-\sqrt{x}}(1+w\cos\sqrt{x})$ for any $w$. For this reason, as far as $\mu_0$ is concerned, we will  treat the measure as unknown, and employ $d\tilde\mu(x)$ only for $\mu_n$, $n\geq1$.

We  separate the integration domain $I\equiv [0,\infty[$ into the origin $0$ and $I_L\equiv ]0,L]$, where eventually we will take  $L\to \infty$. 
Physically this corresponds to separate out the contribution of  states with infinite mass but fixed spin $J\geq1$. As noted already on page \pageref{pagedecrease}, heavy states contribute more to lower, rather than higher, moments; in what follows we will see that in this case infinitely heavy states ($x=0$) contribute only to the first moment.
Indeed, moments $\mu_n$ with $n\geq 1$ have no support on 0 and therefore
 $\mu_n=\mu_n^{I_L}$, where $\mu_n^{I_L}=\int_{I_L}x^nd\mu(x)>0$ are also moments.
  Instead $\mu_0$ has support in 0 (we call it $\mu_0^0\geq 0$), and we can write $\mu_0=\mu_0^0+\mu_0^{I_L}\geq \mu_0^{I_L}$.

To prove that $\mu_0^{I_L}$ can be as small as $1$, we  define a sequence of functions built using positive powers of $x$, $f_n(x)= \sum_{k=1}^n a_k x^k$, so that
\begin{equation}\label{conditions}
0\leq 1-f_n(x)\leq1  \quad x\in I_L\,,
\end{equation}
and such that $f_n$ converges pointwise to unity  $\lim_{n\to\infty}f_n(x)\to 1$ on $I_L$
(for instance, a candidate for $f_n(x)$ is $1/x$ times the  Taylor expansion of $1/x$ in $L$).
Then, integrating over the measure,
\begin{equation}
 0\leq \int_{I_L} (1-f_n(x)) d\mu(x)=\mu_0^{I_L} - \int_{I_L} f_n(x) d\tilde\mu(x)\,,
\end{equation}
where the last integral can be written in terms of higher moments $\sum_{k=1}^na_k\mu_k^{I_L}$ and provides an expression for the minimum of $\mu_0^{I_L}$ in terms of higher-moments only.
Now we can use the dominated convergence theorem to write
\begin{equation}\label{mumin}
\mu_0 \geq \mu_0^L \geq \lim_{n\to\infty}\int_{I_L} f_n(x) d\tilde\mu(x)=\int_{I_L} d\tilde\mu(x)\,.
\end{equation}

Since $x$ corresponds to (the square of) impact parameter, in theories with a mass gap, the measure $\mu(x)$ falls off exponentially at large $x$.

Since the measure decreases exponentially at large $x$, the limit $L\to \infty$ is regular, and the arguments still hold. For $L\to \infty$, the integral in \eq{mumin} is known, and we find, $\mu_0\geq \int d\tilde\mu(x)=2$ which is equivalent to
\be
\tilde{g}_{2,0} \leq \frac{1}{3}\,.
\label{eq:g2c}
\ee
This implies that the kink must lie at $(\tilde{g}_{2,1},\tilde{g}_{2,0}) =({4}/{3},{1}/{3})$, corresponding to a theory of a spin-1 particle at mass $M$, and higher-spin states at $\infty$, as in \eq{rhoinft}.

\vspace{5mm}
\noindent{\bf Higher kinks.} 
\noindent
These  arguments can  be exploited also to obtain the kink positions in the planes of other Wilson coefficients.
In particular, \emph{considering only spins $J\geq1$} (i.e. singling out the spin-0 contribution), one finds that the allowed regions of  $(\tilde g_{n,1},\tilde g_{n,0})$ have also kinks, see Fig.~\ref{g3splot}. Again, numerically the position of such kinks converges very slowly, but analytically we can prove that their position is at,
\begin{equation}\label{higherkink}
(\tilde{g}_{n,1},\tilde{g}_{n,0})=\left(2/3,1/3\right) \,,\qquad \text{for } n>2\ .
\end{equation}
Indeed, from \eq{gns}, we  have for $J\geq1$ that $\frac{g_{n,1}}{g_{n,0}}\geq 2$.
Moving along  the extremal line ${g_{n,1}}=2{g_{n,0}}$, corresponding to a spin-1 state, the kink  resides  at the largest value of $g_{n,0}/g_{1,0}$.
To find this value one has to ask whether it is possible add to the spin-1 state a  spectrum of states  that enhance $g_{n,0}/g_{1,0}$ without affecting the ratio ${g_{n,1}}/{g_{n,0}}$ ($n=3,4,...$). 
This latter condition implies,  from \eq{gns}, $\la \frac{{\cal J}^2}{ m^{2n}}\ra\to 0$
and therefore also $\la {1}/{m^{2n}}\ra\to 0$.
This leads to
\be
\tilde{g}_{n,0} =  \frac{1}{ 1+\frac{M^2}{g^2_{\rho\pi\pi}}\la \frac{1}{ m^2}\rangle}\,.
\ee
Following the same reasoning  as before, one can obtain that 
$\frac{M^2}{g^2_{\rho\pi\pi}}\la \frac{1}{\hat m^2}\rangle\geq 2$, and therefore
$\tilde{g}_{n,0}\leq 1/3$, leading to  \eq{higherkink}. Also this kink lies at the extremum of the spin-1 line.

\section{The $su$-models}
\label{appc}

Let us consider the most general theory of a degenerate spectrum
that contributes to the four-pion amplitude  ${\cal M}(s,u)$  \cite{Caron-Huot:2020cmc,Caron-Huot:2021rmr}.
This means that all states have equal mass $m$, and therefore the denominator of this amplitude
is fixed to be  ${\cal M}(s,u)\propto 1/((s-m^2)(u-m^2))$. 
If we further demand that \eq{UVa} and \eq{UVb} are satisfied for  $k_{\rm min}=1$, we are led to  
\be
{\cal M}(s,u)=\frac{a_1 m^4+a_2 m^2(s+u)+a_3 su}{(s-m^2)(u-m^2)}\,,
\label{suampb}
\ee
where $a_i$ are constants. 
 The Adler's zero condition fixes $a_1=0$. Then, aside from a global multiplicative factor, the amplitude has only one free parameter. We can write it as
\be
{\cal M}^{(su)}_1(s,u)=\frac{m^2(s+u)+\lambda su}{(s-m^2)(u-m^2)}\,,
\label{suamp}
\ee
where the possible values of $\lambda$ are determined by unitarity. Indeed, imposing the positivity of the residues of \eq{suamp}, we obtain
\be\label{lambdarange}
 -2\leq \lambda \leq \frac{2\ln 2-1}{1-\ln2}\,.
 \ee
 In the limiting case $\lambda=-2$, the residues of all   $J>0$ states are zero, and we are left with the scalar amplitude \eq{s}. 
 In the other limit,
 \be
 \lambda=\frac{2\ln2-1}{1-\ln2}\simeq 1.26\,,
 \label{lambda}
 \ee
  the residue of the spin-0 state is zero, leading to an amplitude mediated by an infinite tower of states of spin  $J>0$ and mass $m$.
We will refer to this latter case as the $J>0$ $su$-model.

Expanding \eq{suamp}  for $s,u\ll m^2$, we can obtain the  Wilson coefficients:
\be
g_{n,0}= \frac{1}{m^{2n}}\ , \qquad  g_{n,l}= \frac{2+\lambda}{m^{2n}}\ \ \ (n,l>0) \,.
\ee
For  \eq{lambda}, the Wilson coefficients, normalized as in \eq{ratioW} for $M=m$, are given by
\be
\tilde{g}_{n,0}= 1\ , \qquad  \tilde{g}_{n,l} =\frac{1}{1-\ln2}\simeq 3.26\ \ \ (n>1,l>0) \,.
\label{suwilsons}
\ee
These are shown in Figs.~\ref{fig:g2g2pdec}--\ref{g3splot} as a black dot.

We can proceed in a similar way to construct the most general amplitude of degenerate high-spin states, but now satisfying \eq{UVa} and \eq{UVb} for $k_{\rm min}=2$. Imposing the Adler's zero condition, such amplitude is given by (up to a multiplicative constant factor)
\be
{\cal M}^{(su)}_2(s,u)=\frac{m^2(s+u)+\lambda_1 su + \lambda_2 (s^2+u^2)}{(s-m^2)(u-m^2)}\,.
\label{suamp2}
\ee
The values of the constants  $\lambda_1$ and $\lambda_2$  determine different theories according to:
\begin{itemize}
	\item For $2+\lambda_1+2\lambda_2=0$, the residues for $J>1$ vanish and we have a theory with only a spin-0 and spin-1 state.
	
	\item For $\lambda_2=(2+\lambda_1)\frac{6\ln8-12}{25-36\ln2}$, the residue for the spin-1 state is zero and we have a theory with $J=0$ and $J>1$.
	
	\item For $\lambda_2=-1 + \frac{1 - \lambda_1 ( \ln4-2)}{ \ln16-1}$, the residue for the spin-0 vanishes and we have a theory with spins $J>0$.
\end{itemize}
Furthermore, for  $\lambda_1=-2$, $\lambda_2=0$, we recover the scalar amplitude \eq{s}, while for
$\lambda_1=\lambda_2=-{2}/{3}$ we obtain  the  spin-1 amplitude \eq{rho}. 
We are interested in the case where the amplitude \eq{suamp2} contains only
$J>1$ states which corresponds to taking
\be
\lambda_1=\frac{20\ln2-13}{19-28\ln2}\ , \ \ \ 
\lambda_2=\frac{6\ln8-12}{19-28\ln2}\,.
\label{lambdas}
\ee
We will refer to this  case as the $J>1$ $su$-model.

The Wilson coefficients of \eq{suamp2} are given by
\be
\begin{split}
g_{1,0}= \frac{1}{m^{2}}\ ,\qquad g_{2,1}= \frac{2+\lambda_1}{m^{4}}\ ,\qquad g_{n,0}= \frac{1+\lambda_2}{m^{2n}}\ \ \ (n>1)\,, \\
g_{n,1}= \frac{2+\lambda_1+\lambda_2}{m^{2n}} \ , \qquad  g_{n,l}= \frac{2+\lambda_1+2\lambda_2}{m^{2n}}
\ \ \ (n,l>1) \,.
\end{split}
\ee
For the case \eq{lambdas}, normalizing the coefficients as given in \eq{ratioW2}, we have 

\be
\bar{g}_{n,0}= 1\ , \ \ \  \bar{g}_{n,1}= \frac{18\ln 2-13}{10\ln 2-7}\simeq 7.64\ , \ \ \   \bar{g}_{n,l}= \frac{1}{7-10\ln 2}\simeq 14.59\ \ \ (n>2,l>1) \,.
\label{su-s-v}
\ee
These are shown in Fig.~\ref{g34s} as a black dot.

\subsection{Two-mass $su$-model}

The $su$-models define part of the boundaries of the allowed regions of the Wilson coefficients.
To see this, we can deform the above $su$-models by introducing an additional  pole in the amplitude, i.e.,
${\cal M}(s,u)\propto 1/((s-m^2)(u-m^2)(s-M^2)(u-M^2))$.
In this case the most general amplitude can be written as
\be
{\cal M}(s,u) ={\cal M}^{(su)}_1(m)+\alpha{\cal M}^{(su)}_1(M)+
\frac{\beta\, s^2u^2}{(s-m^2)(u-m^2)(s-M^2)(u-M^2)}\ ,
\label{sudeform}
\ee
that corresponds to  two $su$-models  (\eq{suamp}) with mass $m$ and $M$ respectively, and an extra term.
Apart from the masses,  the amplitude has 4 parameters: the two $\lambda$ of the $su$-models,
$\alpha$ and $\beta$. 
We are interested in this model without the scalars. 
Removing the scalars in the two $su$-models
fixes the $\lambda$'s  to the value  \eq{lambda}.
Removing the scalar from the last term of \eq{sudeform} corresponds to adding  to the amplitude the term
\be
\beta\left[f(m,M)\left(\frac{1}{s-m^2}+\frac{1}{u-m^2}\right)+(M\leftrightarrow m)\right]\,,
\label{subsudeform}
\ee
where
\be
f(m,M)=\frac{m^4M^2+m^6\left(\ln 2-1\right)+m^2M^4 \ln \frac{M^2}{m^2+M^2}}{\left(m^2-M^2\right)^2}\,.
\ee
Requiring the positivity of the spectral function for the $J>0$ states in \eq{sudeform}
leads to $\beta\geq 0$.

\eq{sudeform} with \eq{subsudeform} leads to 
\be
\frac{\tilde{g}_{2,1}}{\tilde{g}_{2,0}} = \frac{3.26\left(\frac{1}{m^4} + \frac{a}{M^4}\right)}
{\left(\frac{1}{m^4} + \frac{a}{M^4}\right)-\beta(\frac{f(m,M)}{m^4}+\frac{f(M,m)}{M^4})}
\,.
\label{ratiosu2}
\ee
Since $\beta(\frac{f(m,M)}{m^4}+\frac{f(M,m)}{M^4})$ is a positive-definite function,
we see that the ratio ${\tilde{g}_{2,1}}/{\tilde{g}_{2,0}}$
is bounded  from below by the $su$-model.

\section{The Lovelace-Shapiro amplitude}
\label{appd}

The Lovelace-Shapiro (LS) amplitude for the scattering of four pions is defined as \cite{Lovelace:1968kjy,Shapiro:1969km}
\be
{\cal M}^{(\text{LS})}(s,u)=\frac{\Gamma(1-\alpha(s))\Gamma(1-\alpha(u))}{\Gamma(1-\alpha(s)-\alpha(u))}\ ,
\label{LSamp0}
\ee
where $\alpha(s)=\alpha_0+\alpha' s$ is referred as the Regge trajectory.
We will fix the  values of $\alpha_0$ and $\alpha'$  by requiring that \eq{LSamp} satisfies the Adler zero condition,
${\cal M}^{(\text{LS})}(s,u)\to 0$ for $s,u\to 0$, and that the first pole of  \eq{LSamp}  occurs for 
$s=m_\rho^2$.
These two conditions lead to $\alpha_0 = 1/2$ and $\alpha'=1/(2m_\rho^2)$ \cite{Bianchi:2020cfc} and then we can write
\be
{\cal M}^{(\text{LS})}(s,u)=\frac{\Gamma\left(\frac{1}{2}-\frac{s}{2m^2_\rho}\right)
\Gamma\left(\frac{1}{2}-\frac{u}{2m^2_\rho}\right)}
{\Gamma\left(\frac{t}{2m^2_\rho}\right)}\ .
\label{LSamp}
\ee
By looking at the poles of  \eq{LSamp}, 
one can see that the LS amplitude corresponds to a theory of higher-spin states with  masses 
\be
m_n^2= m_\rho^2(2n+1) \,,\ \  n=0,1,2,...\,.
\ee
For a given $n$, there are at most $n + 1$ states with  spin  $J = 0, 1, ..., n + 1$.
Furthermore, \eq{LSamp}  satisfies the condition \eq{UVa} and \eq{UVb} with $k_{\rm min}=1$.

The first Wilson coefficients arising from  \eq{LSamp} in a low-energy expansion are given by
\bea
	g_{1,0}&=&\frac{\pi}{2m_\rho^2}\ ,\qquad  
	g_{2,0} = \frac{1}{2}g_{2,1}= \frac{\pi\ln 2}{2m_\rho^4}\ ,\qquad 
	g_{3,0}=\frac{\pi^3+12 \pi\ln^2 2}{48 m_\rho^6}\ ,\\ 
	g_{3,1}&=&\frac{3\pi\ln^2 2}{4 m_\rho^6}\ ,\qquad
	g_{4,0}=\frac{\pi\left(\pi^2 \ln 2+4\ln^3 2+6\zeta(3)\right)}{48 m_\rho^8}\ ,\\ 	
	g_{4,1}&=&\frac{\pi\left(\pi^2\ln 2+16\ln^3 2+3\zeta(3)\right)}{48 m_\rho^8}\ ,\qquad
	g_{4,2}=\frac{\pi\left(4\ln^3 2-\zeta(3)\right)}{8 m_\rho^8}\,,
\eea
with $\zeta$ the Riemann zeta function. 
For the  normalized  Wilson coefficients defined in \eq{ratioW} we have, taking $M=m_\rho$,
\be
	\tilde{g}_{2,0}\simeq 0.69\,,\  
	\tilde{g}_{2,1}\simeq 1.39\,,\ 
	\tilde{g}_{3,0}\simeq 0.65\,,\ 
	 \tilde{g}_{3,1}\simeq 0.72\,,\
	\tilde{g}_{4,0}\simeq 0.64\,,\ 
	\tilde{g}_{4,1}\simeq 0.66\,,\
	\tilde{g}_{4,2}\simeq 0.03\,,
\ee
while for the normalized coefficients in \eq{ratioW2}, we have
\be
	\bar{g}_{2,1} \simeq 2\,,\
	\bar{g}_{3,0}\simeq0.94\,,\
	\bar{g}_{3,1}\simeq1.04\,,\
	\bar{g}_{4,0}\simeq0.92\,,\
 	\bar{g}_{4,1}\simeq0.95\,,\
	\bar{g}_{4,2}\simeq0.05\,.
\ee
Since a theory of scalars provides a consistent UV completion of the pion amplitude ${\cal M}$, 
satisfying \eq{UVa} and \eq{UVb} with $k_{\rm min}=1$,
we can find a new  consistent amplitude by subtracting the scalars from \eq{LSamp}.
This leads to
\be
{\cal M}^{(\text{LS})}_{J>0}(s,u)={\cal M}^{(\text{LS})}(s,u) -\sum_{n=0}^\infty \left[\frac{m_n^2}{s-m_n^2}\kappa_{s,0}^{\text{LS}}+(s\leftrightarrow u)\right]\,,
\label{LS-Samp}
\ee
with
\be
\kappa_{s,J}^{\text{LS}}=\frac{2J+1}{2}\int_{-1}^1 dx \ P_J(x)\underset{s=m_n^2}{\text{Res}}\left[{\cal M}^{(\text{LS})}(s,u(x))\right]\,,
\ee
where $u(x)=-s(1-x)/2$. 
From \eq{LS-Samp}, we obtain  by expanding at small $s,u$:
\be
\tilde{g}_{2,0}\simeq 0.55\,,\
\tilde{g}_{2,1}\simeq 2.05\,,\
\tilde{g}_{3,0}\simeq 0.49\,,\
\tilde{g}_{3,1}\simeq 1.07\,,\
\tilde{g}_{4,0}\simeq 0.48\,,\
\tilde{g}_{4,1}\simeq 0.97\,,\
\tilde{g}_{4,2}\simeq 0.05\,.
\label{LSwilsons}
\ee
In a similar way, we can also remove the (infinite) spin-1 states of \eq{LS-Samp} to obtain an amplitude that still satisfies the Froissart-Martin condition, \eq{UVa} and \eq{UVb} with $k_{\rm min}=2$:
\be
{\cal M}^{(\text{LS})}_{J>1}(s,u)={\cal M}^{(\text{LS})}_{J>0}(s,u) -\sum_{n=0}^\infty \left[\frac{m_n^2+2u}{s-m_n^2}\kappa_{s,1}^{\text{LS}}+(s\leftrightarrow u)\right]\ .
\label{LS-S-1amp}
\ee
From \eq{LS-S-1amp} we obtain
\be
		\bar{g}_{2,1}\simeq 0.97\,,\
		  \bar{g}_{3,0}\simeq 0.71\,,\
		  \bar{g}_{3,1}\simeq 5.99\ ,\
		\bar{g}_{4,0}\simeq 0.60\,,\
		\bar{g}_{4,1}\simeq 4.18\,,\
		\bar{g}_{4,2}\simeq 8.15\,.
\ee

\section{The Coon amplitude}
\label{appe}

The Lovelace-Shapiro amplitude presented in Appendix~\ref{appd} can be generalized to a larger class of amplitudes depending on an additional parameter $q$. This is the so-called Coon amplitude, which was first proposed in \cite{Coon:1972qz}\footnote{The idea of the Coon amplitude goes back to an earlier work by Coon \cite{Coon:1969yw}, where he defined a generalization of the Veneziano amplitude which was slightly different from \eq{coonamp}. Shortly after that he proposed the Coon version of Lovelace-Shapiro amplitude together with Sukhatme and Tran Thanh Van in \cite{Coon:1972qz}.}:
\be
{\cal M}_q(s,u)= C(\sigma,\tau,q)\prod_{n=0}^\infty\frac{\left(1-q^{n+1}\right)\left(\sigma\tau-q^{n+1}\right)}{\left(\sigma-q^{n+1}\right)\left(\tau-q^{n+1}\right)}\ ,
\label{coonamp}
\ee
where $\sigma=1+(q-1)(\alpha_0+\alpha' s)$ and  $\tau=1+(q-1)(\alpha_0+\alpha' u)$. As explained in Appendix~\ref{appd}, we take $\alpha_0 = 1/2$ and $\alpha'=1/(2m_\rho^2)$. The parameter $q$ takes values between 0 and 1, and in the limit $q\rightarrow 1$ we recover the LS amplitude \eq{LSamp}. There is some freedom in the choice of the prefactor $C$, as long as it satisfies $\lim_{q\rightarrow 1}C(\sigma,\tau,q)=1$.

The Coon amplitude has an infinite number of simple poles at
\be
s_n=m_\rho^2\frac{1+q-2q^{n+1}}{1-q}\ ,\qquad n=0,1,2,...\ .
\ee
The corresponding residues are
\be
\underset{s=s_n}{\text{Res}}\ {\cal M}_q(s,u)=C(\sigma_n,\tau,q)\  \frac{2 q^{n+1}}{1-q}\ \frac{\tau-1}{\tau^{n+1}}\ m_\rho^2\prod_{l=0}^{n-1}\frac{(\tau-q^{l-n})}{(1-q^{l-n})}\ ,
\label{rescoon}
\ee
where $\sigma_n=\sigma (s=s_n)$. 
It is important to remark  that the spectrum  has an \textit{accumulation point} at $s_*=\lim_{n\rightarrow\infty}s_n=m_\rho^2\frac{1+q}{1-q}$. In the limit $q\rightarrow 1$, the accumulation point is located at infinity and we recover the evenly-spaced spectrum of the LS amplitude.

It is customary to fix the prefactor $C(\sigma,\tau,q)$ with the further assumption that the residues of the Coon amplitude are polynomials in $u$, since it is believed that non-polynomial residues lead to problems with the locality of the theory. The prefactor is in this case set to
\be
C(\sigma,\tau,q)=q^{\frac{\ln\sigma\ln\tau}{\ln q\ln q}}\ ,
\label{prefcoon}
\ee
which reduces to $C(\sigma_n,\tau,q)=\tau^{n+1}$ at the $s_n$ pole. This term cancels the factor $\tau^{n+1}$ in the denominator of \eq{rescoon} and ensures that the residues are polynomials. 
In this case,  we have  that for any $n$, there are $n+1$ states with spin $J=0,1,...,n+1$,  as in the LS amplitude.

Using the prefactor \eq{prefcoon}  makes however the Coon amplitude \eq{coonamp} non-meromorphic. In addition to the simple poles, there is a branch cut starting at the accumulation point $s_*$. Although the physical meaning of this kind of singularities is unclear, amplitudes with branch cuts can still obey the requirements of unitarity, crossing symmetry and Regge boundedness, so it is interesting to include them in our study.

Regarding the high-energy behavior, the amplitude with prefactor \eq{prefcoon} grows at fixed $u$ like ${\cal M}_q(s,u)\sim f(u)\ s^{\ln \tau/\ln q} $. For negative $u$, $\ln \tau/\ln q< 0.5$, so the amplitude obeys \eq{UVa} for $k_{\rm min}=1$. At fixed $t$, the amplitude grows like ${\cal M}_q(s,u)\sim s^{\ln ((1-q)\alpha' s)/\ln q} $. Since $\ln ((1-q)\alpha' s)/\ln q<0.5$, it also obeys \eq{UVb} for $k_{\rm min}=1$.

The last point to address is the unitarity of the Coon amplitude. We have found that the $J=0$ states have negative residues for all $n>0$, making the amplitude non-unitary. This result is in agreement with the early study \cite{Coon:1974iu}\footnote{Some recent works \cite{Figueroa:2022onw,Geiser:2022icl,Chakravarty:2022vrp} on the Veneziano version of the Coon amplitude have shown that it is unitary in $D=4$ dimensions. Our result agrees with the results of Coon and Yu \cite{Coon:1974iu}, who considered the Lovelace-Shapiro version of the Coon amplitude.}, which pointed out the presence of ghosts. On the other hand, we have found that the rest of the $J$ states  have positive residues inside the scope of our numerical searches. 
As for the branch cut discontinuity, it can also be expanded in partial waves according to \eq{pw}. We have obtained that the spectral density $\rho_J(s)$ is positive for all $J$ except from $J=0$.
Due to this problems, we will not discuss this case further.


On the other hand, relaxing the assumption of polynomial residues, we can simply take $C(\sigma,\tau,q)=1$, 
and  the amplitude becomes meromorphic (there is no branch cut).
We now have  an infinite number of spins  present for each $n$. 
At   large $s$, taking both fixed $u$ and $t$, we have  ${\cal M}_q(s,u)\sim {\text constant}$. 
 Therefore  it satisfies both \eq{UVa} and \eq{UVb} for  $k_{\rm min}=1$.
For $C(\sigma,\tau,q)=1$ we have also found that  the only negative residue corresponds to the $J=0$, $n=1$ state only in the small region $0.94\lesssim q< 1$.

The contribution of the Coon amplitude  with $C=1$  is shown in Fig.~\ref{fig:g2g2pdec}
for $q\in [0,1]$. We also show the case in which all scalars are removed from the spectrum.
In the limit $q\rightarrow 1$ the Coon amplitude approaches the $J>0$ Lovelace-Shapiro model, while for $q\rightarrow 1$, the Coon amplitude is dominated by the 
 $n=0$ level (the  $n>0$ levels decouple), giving the $J>0$ $su$-model.
\bibliography{draft}
\bibliographystyle{apsrev} 

\end{document}